\documentclass[a4paper,11pt]{article}
\pdfoutput=1

\usepackage{jheppub}

\usepackage{subfig}
\usepackage{xspace}
\usepackage[countmax]{subfloat}
\usepackage{slashed}
\usepackage{longtable}
\newcommand*{\thead}[1]{\multicolumn{1}{c}{\bfseries #1}}
\usepackage{booktabs}
\usepackage{comment}

\usepackage{bbm}

\def\be{\begin{equation}}
\def\ee{\end{equation}}

\DeclareRobustCommand{\Sec}[1]{Sec.~\ref{#1}}

\DeclareRobustCommand{\App}[1]{App.~\ref{#1}}

\DeclareRobustCommand{\Fig}[1]{Fig.~\ref{#1}}
\DeclareRobustCommand{\Figs}[2]{Figs.~\ref{#1} and \ref{#2}}
\DeclareRobustCommand{\Eq}[1]{Eq.~(\ref{#1})}

\DeclareRobustCommand{\Ref}[1]{Ref.~\cite{#1}}

\title{How Much Information is in a Jet?}

\author{Kaustuv Datta and Andrew Larkoski}
\affiliation{Physics Department, Reed College, Portland, OR 97202, USA}

\emailAdd{dattak@reed.edu}
\emailAdd{larkoski@reed.edu}

\abstract{
Machine learning techniques are increasingly being applied toward data analyses at the Large Hadron Collider, especially with applications for discrimination of jets with different originating particles.  Previous studies of the power of machine learning to jet physics have typically employed image recognition, natural language processing, or other algorithms that have been extensively developed in computer science.  While these studies have demonstrated impressive discrimination power, often exceeding that of widely-used observables, they have been formulated in a non-constructive manner and it is not clear what additional information the machines are learning.  In this paper, we study machine learning for jet physics constructively, expressing all of the information in a jet onto sets of observables that completely and minimally span $N$-body phase space.  For concreteness, we study the application of machine learning for discrimination of boosted, hadronic decays of $Z$ bosons from jets initiated by QCD processes.  Our results demonstrate that the information in a jet that is useful for discrimination power of QCD jets from $Z$ bosons is saturated by only considering observables that are sensitive to 4-body (8 dimensional) phase space.
}

\begin{document} 
\maketitle

\section{Introduction}\label{sec:intro}

The problem of discrimination and identification of high energy jet-like objects observed at the Large Hadron Collider (LHC) is fundamental for both Standard Model physics and searches as the lower bound on new physics mass scales increase.  Heavy particles of the Standard Model, like the $W$, $Z$, and $H$ bosons or the top quark, can be produced with large Lorentz boosts and dominantly decay through hadrons.  They will therefore appear collimated in the detector and similar to that of jets initiated by light QCD partons.  The past several years have seen a huge number of observables and techniques devoted to jet identification \cite{Adams:2015hiv,Altheimer:2013yza,Altheimer:2012mn,Abdesselam:2010pt}, and many have become standard tools in the ATLAS and CMS experiments.

The list of observables for jet discrimination is a bit dizzying, and in many cases there is no organizing principle for which observables work well in what situations.\footnote{There has been some effort in the past to identify and quantify (over)complete bases of jet observables \cite{Tkachov:1995kk,Sveshnikov:1995vi,Cherzor:1997ak,Tkachov:1999py}.}  Motivated by the large number of variables that define the structure of a jet, several groups have recently applied machine learning methods to the problem of jet identification \cite{Cogan:2014oua,Almeida:2015jua,deOliveira:2015xxd,Baldi:2016fql,Guest:2016iqz,Conway:2016caq,Barnard:2016qma,Komiske:2016rsd,deOliveira:2017pjk,Kasieczka:2017nvn,Louppe:2017ipp,Dery:2017fap,Pearkes:2017hku}.  Rather than developing clever observables that identify certain physics aspects of the jets, the idea of the machine learning approach is to have a computer construct an approximation to the optimal classifier that discriminates signal from background.  For example, \Ref{deOliveira:2015xxd} interpreted the jet detected by the calorimetry as an image, with the pixels corresponding to the calorimeter cells and the ``color'' of the pixel corresponding to the deposited transverse momentum in the cell.  These techniques have outperformed standard jet discrimination observables and show that there is additional information in jets to exploit.

However, this comes with a significant cost.  Machine learning methods applied to jet physics typically have hundreds of input variables with thousands of correlations between them.  Thus, in one sense this problem seems ideally suited for machine learning, but it also lacks the immediate physical interpretation and intuition that individual observables have.  Previous studies have shown that the computer is learning information about what discriminates jets of different origins, but it has not been clearly demonstrated what information standard observables are missing.  Along these same lines, the improvement of discrimination performance of machine learning over standard observables is relatively small, suggesting that standard observables capture the vast majority of useful information in jets.

In this paper, we approach machine learning for jet discrimination from a different perspective.  We construct an observable basis that completely and minimally spans the phase space for the substructure of a jet.\footnote{By ``span'' we do not mean in the vector space sense.  Rather, the measurement of the basis of observables defines a system of equations that can be inverted to uniquely determine the phase space variables.}  For a jet with $M$ particles, the phase space is $3M-4$ dimensional, and so we identify $3M-4$ infrared and collinear (IRC) safe jet substructure observables that span the phase space.\footnote{Note that this will completely define the phase space of the jet substructure; that is the relative configuration of emissions in the jet.  It will not identify the total jet $(p_T,\eta,\phi)$.  This may be useful information, but is explicitly sensitive to global event properties which is beyond the scope of this paper.  We thank Ben Nachman for emphasizing this point.}  These basis observables are then passed to a machine learning algorithm for identification of relevant discrimination information.\footnote{Because we input a finite number of IRC safe observables to the machine, its output classifier will in general be Sudakov safe \cite{Larkoski:2013paa,Larkoski:2015lea}.}  A general jet will have an arbitrary number of particles in it, and so we will observe how the discrimination power depends on the dimension of phase space that we assume.  That is, we will assume that the jet has 2 particles, 3 particles, 4 particles, etc., as defined by the set of basis observables and observe how the discrimination power improves.  This method is constructive in the following sense.  With some number of assumed particles in the jet, the discrimination power will saturate, which then immediately tells us what reduced set of observables are necessary to effectively extract all information that is useful for discrimination.  This approach has the additional advantage that the identified observables can be calculated theoretically from first principles, without relying on parton shower modeling.

\begin{figure}
\begin{center}
\includegraphics[width=.6\textwidth]{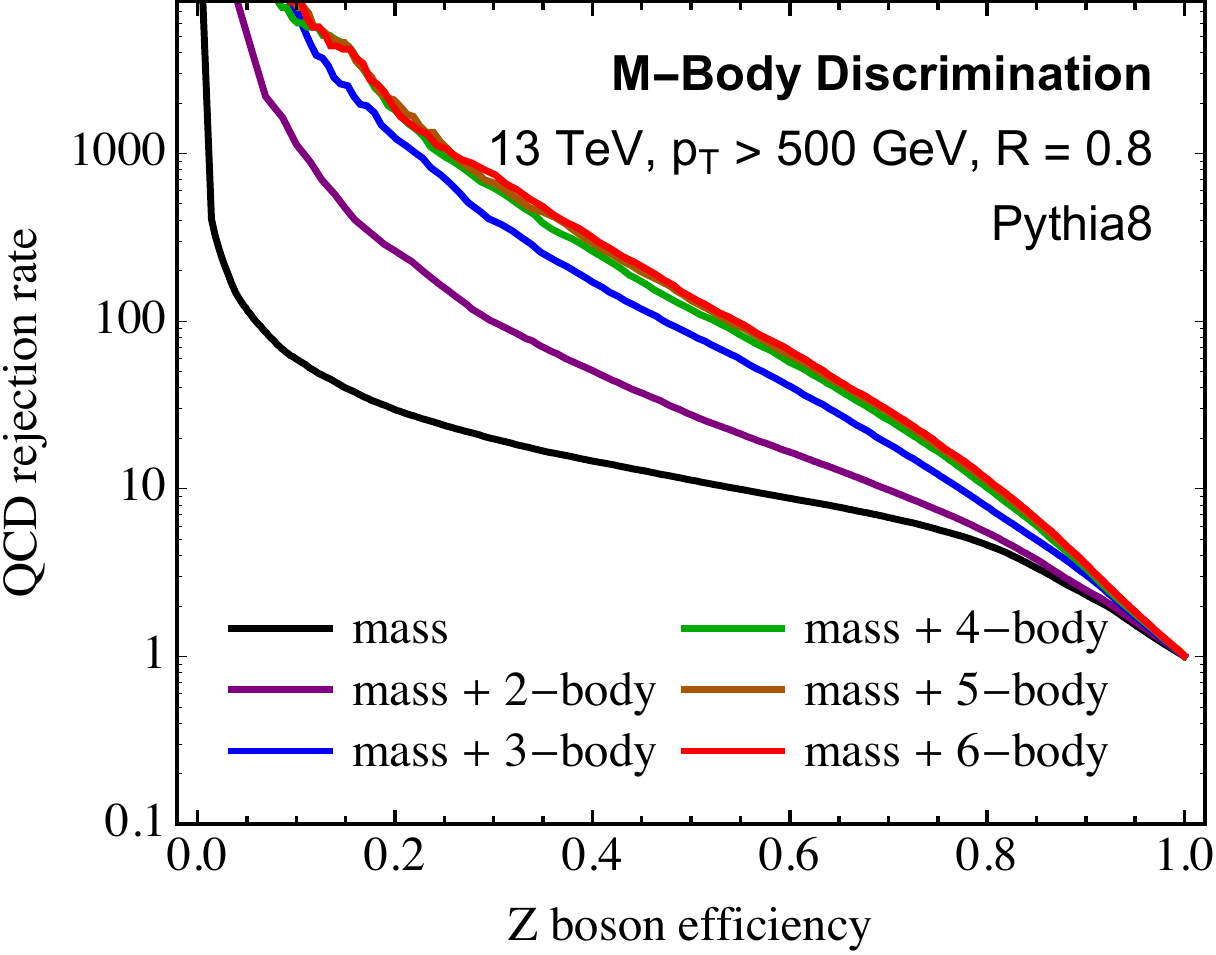}
\caption{
$Z$ boson jet efficiency vs.~QCD jet rejection rate plot as generated by the deep neural network. 
Details of the event simulation, jet finding, and machine learning are described in \Sec{sec:deeplearn}.  The different curves correspond to the mass plus collections of observables that uniquely define $M$-body phase space.  Discrimination power is seen to saturate when 4-body phase space is resolved.
}
\label{fig:introplot}
\end{center}
\end{figure}

As it is a widely-studied problem in jet substructure, we will apply this approach to the discrimination of boosted, hadronically decaying $Z$ bosons from jets initiated by light quarks or gluons.  The results of our study are shown in \Fig{fig:introplot}.  Here, we plot the simulated signal ($Z$ boson) efficiency versus the background (QCD jet) rejection rate as determined by a deep neural network, for observables that are sensitive to 2-, 3-, 4-, 5- and 6-body phase space.  To identify the phase space variables, we choose to measure the jet mass and the $N$-subjettiness observables \cite{Stewart:2010tn,Thaler:2010tr,Thaler:2011gf}, but this choice is not special.  This plot demonstrates that observables sensitive to 4-body phase space saturate the discrimination power.  4-body phase space is only 8 dimensional, suggesting that very few observables are necessary to identify all interesting structure of these jets.  We anticipate that this approach can be applied to other discrimination problems in jet substructure, as well, and greatly reduce the dimensionality of the variable space that is being studied.

The outline of this paper is as follows.  In \Sec{sec:basis}, we define the observable basis that is used to identify all variables of $M$-body phase space.  As mentioned above, we choose to use the $N$-subjettiness observables.  In this section, we also prove that the set of observables is complete and minimal.  In \Sec{sec:deeplearn}, we discuss our event simulation and machine learning implementation.  We present the results of our study, and compare discrimination power from the $M$-body phase space observables to standard observables as a benchmark.  We conclude in \Sec{sec:conc}.  Additional details are in the appendices.

\section{Observable Basis}\label{sec:basis}

In this section, we specify the basis of IRC safe observables that we use to identify structure in the jet.  For simplicity, we will exclusively use the $N$-subjettiness observables \cite{Stewart:2010tn,Thaler:2010tr,Thaler:2011gf}, however this choice is not special.  One could equivalently use the originally-defined $N$-point energy correlation functions \cite{Larkoski:2013eya}, or their generalization to different angular dependence \cite{Moult:2016cvt}.  Our choice of using the $N$-subjettiness observbles in this analysis is mostly practical: the evaluation time for the $N$-subjettiness observables is significantly less than for the energy correlation functions.  We also emphasize that the particular choice of observables below is to just ensure that they actually span the phase space for emissions in a jet.  There may be a more optimal choice of a basis of observables, but optimization of the basis is beyond this paper.

The $N$-subjettiness observable $\tau_N^{(\beta)}$ is a measure of the radiation about $N$ axes in the jet, specified by an angular exponent $\beta>0$:
\begin{equation}
\tau_N^{(\beta)} = \frac{1}{p_{T J}} \sum_{i\in \text{Jet}} p_{Ti} \min\left\{
R_{1i}^\beta,R_{2i}^\beta,\dotsc,R_{Ni}^\beta
\right\}\,.
\end{equation}
In this expression, $p_{TJ}$ is the transverse momentum of the jet of interest, $p_{Ti}$ is the transverse momentum of particle $i$ in the jet, and $R_{Ki}$, for $K=1,2,\dotsc,N$, is the angle in pseudorapidity and azimuth between particle $i$ and axis $K$ in the jet.  There are numerous possible choices for the $N$ axes in the jet; in our numerical implementation, we choose to define them according to the exclusive $k_T$ algorithm \cite{Catani:1993hr,Ellis:1993tq} with standard $E$-scheme recombination \cite{Blazey:2000qt}.  Note that $\tau_N^{(\beta)} = 0$ for a jet with $N$ or fewer particles in it.

To identify structure in the jet, we need to measure an appropriate number of different $N$-subjettiness observables.  This requires an organizing principle to ensure that the basis of observables is complete and minimal.  Our approach to ensuring this is to identify the set of $N$-subjettiness observables that can completely specify the coordinates of $M$-body phase space.  Ensuring that the set is minimal is then straightforward: as $M$-body phase space is $3M-4$ dimensional, we only measure $3M-4$ $N$-subjettiness observables.  A jet also has an overall energy scale.  To ensure sensitivity to this energy scale, we will also measure the jet mass, $m_J$.

\begin{figure}
\begin{center}
\subfloat[]{
\label{fig:2body}
 \includegraphics[width=.27\textwidth]{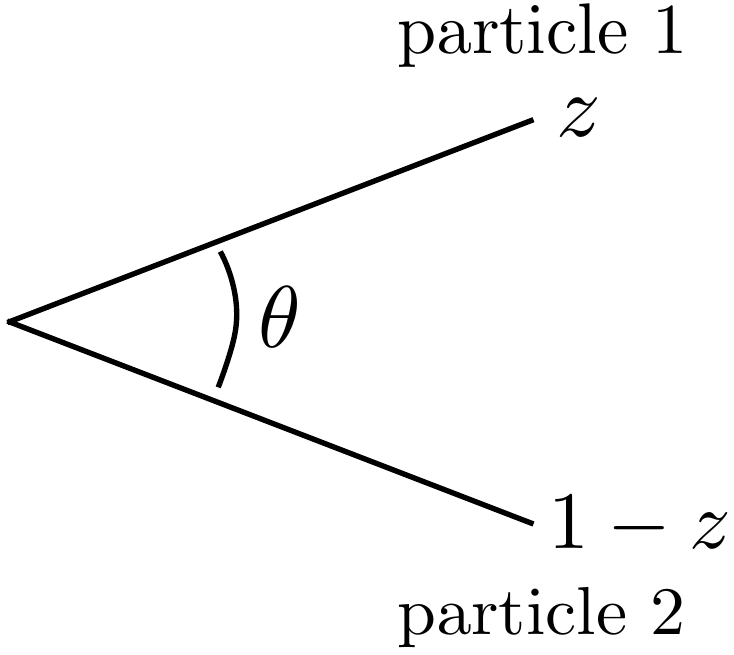}
}
\hspace{2cm}
\subfloat[]{
\label{fig:3body}
 \includegraphics[width=.35\textwidth]{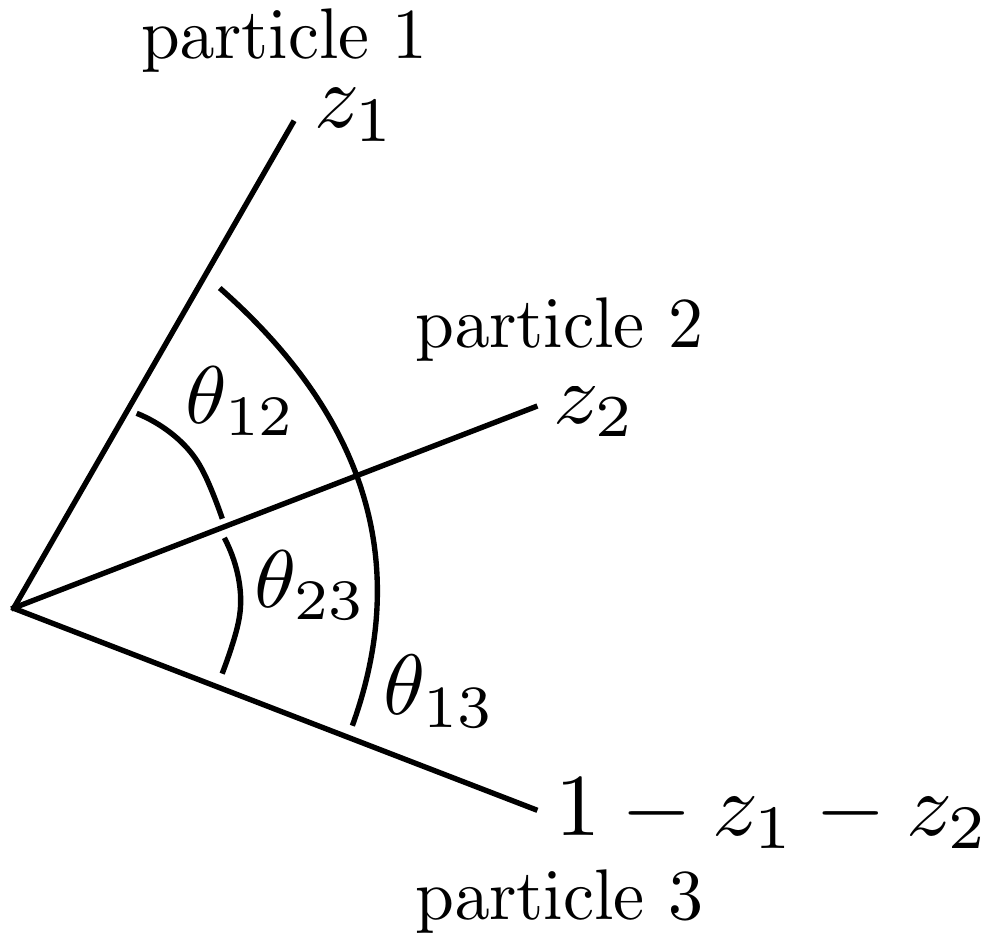}
}
\caption{
Illustration of the momentum fraction and pairwise angle variables that describe 2-body (right) and 3-body (left) phase space.
}
\label{fig:spaces}
\end{center}
\end{figure}

We will describe how to do this for low dimensional phase space, and then generalize to arbitrary $M$-body phase space.  We will work in the limit where the jet is narrow and so all particles in the jet can be considered as relatively collinear.  This simplifies the expressions for the values of the $N$-subjettiness observables to illustrate their content, but does not affect their ability to span the phase space variables.
\begin{itemize}

\item {\bf 2-Body Phase Space:} 2-body phase space is $3\cdot2-4=2$ dimensional.  For a jet with two particles, the phase space can be completely specified by the transverse momentum fraction $z$ of one of the particles:
\begin{align}
&z=\frac{p_{T1}}{p_{TJ}}\,, &1-z=\frac{p_{T2}}{p_{TJ}}\,,
\end{align}
and the splitting angle $\theta$ between the particles.  This configuration is shown in \Fig{fig:2body}.  To uniquely identify the $z$ and $\theta$ of this jet, we can measure two 1-subjettiness observables, defined by different angular exponents $\alpha \neq \beta$.  For concreteness, we will measure $\tau_1^{(1)}$ and $\tau_1^{(2)}$.

To determine the measured values of the $1$-subjettiness observables, we need to determine the angle between the individual particles of the jet and the axis.  Because $E$-scheme recombination conserves momentum, the angles between the particles 1 and 2 and the axis are:
\begin{align}
&\theta_1 = (1-z)\theta\,, &\theta_2 = z\theta\,.
\end{align}
It then follows that the values of the 1-subjettiness observables are:
\begin{align}\label{eq:2bodynsub}
&\tau_1^{(1)} = 2z(1-z)\theta\,, & \tau_1^{(2)} = z(1-z)\theta^2\,.
\end{align}
These expressions can be inverted to find $z$ and $\theta$ individually:
\begin{align}\label{eq:2bodysol}
&z(1-z)=\frac{\left(\tau_1^{(1)}  \right)^2}{4\tau_1^{(2)} } \,, &\theta= 
\frac{2\tau_1^{(2)}}{\tau_1^{(1)}}\,.
\end{align}
Note the symmetry for $z\leftrightarrow 1-z$: this is to be expected because we have not assumed an ordering of the transverse momenta of particles 1 and 2.

\item {\bf 3-Body Phase Space:}  3-body phase space is $3\cdot 3-4=5$ dimensional, and so to completely determine the configuration of a jet with three particles, we need to measure 5 $N$-subjettiness observables.  The 5 phase space variables can be defined to be the 3 pairwise angles between the particles $i$ and $j$ in the jet: $\theta_{12}$, $\theta_{13}$, and $\theta_{23}$, and two of the transverse momentum fractions, say, $z_1$ and $z_2$.  We define the momentum fractions as: 
\begin{align}
z_1=\frac{p_{T1}}{p_{TJ}}\,, \qquad z_2=\frac{p_{T2}}{p_{TJ}}\,, \qquad1-z_1-z_2 = \frac{p_{T3}}{p_{TJ}}\,.
\end{align}
This configuration is shown in \Fig{fig:3body}.  To determine the phase space variables, we will measure a collection of 1- and 2-subjettiness observables.

Our choice for which collection of 1- and 2-subjettiness observables is the following.  We will measure three 1-subjettiness observables $\tau_1^{(0.5)}$, $\tau_1^{(1)}$, and $\tau_1^{(2)}$ and two 2-subjettiness observables $\tau_2^{(1)}$ and $\tau_2^{(2)}$.  To motivate this collection of observables, note that one of the axes for measuring 2-subjettiness necessarily lies along the direction of a particle.  Therefore, measuring 2-subjettiness is only sensitive to one relative energy fraction and one angle between pairs of particles, as illustrated explicitly in the 2-body case in \Eq{eq:2bodynsub}.  Because 2-subjettiness is only sensitive to two phase space variables, we only measure two 2-subjettiness observables.

The axis for the 1-subjettiness observables, however, is necessarily displaced from the direction of any particle in the jet.\footnote{This is an important point, and the reason why we use $E$-scheme recombination as opposed to winner-take-all (WTA) recombination \cite{Bertolini:2013iqa,Larkoski:2014uqa,Larkoski:2014bia}, for example, to define the $N$-subjettiness axes.  Because the axes defined by the WTA scheme necessarily lie along the direction of particles, there are non-degenerate configurations of particles for which measuring 5 $N$-subjettiness observables do not span the full 3-body phase space.}  This is because the $E$-scheme recombination conserves momentum, and so this axis can only degenerate to the direction of a particle in the jet if another particle has 0 energy or is exactly collinear to another particle.  Therefore, this collection of 5 $N$-subjettiness observables will generically span the full 3-body phase space.  In \App{app:3body}, we present the explicit expressions for the 1- and 2-subjettiness observables in terms of the phase space coordinates.

\item {\bf M-Body Phase Space:}  For $M$-body phase space, we can define the coordinates of that phase space by $M-1$ transverse momentum fractions $z_i$, for $i=1,\dotsc,M-1$, and $2M-3$ pairwise angles $\theta_{ij}$ between particles $i$ and $j$.  The remaining 
$$
{M \choose 2} - (2M-3) = \frac{1}{2}(M-2)(M-3)\,,
$$
pairwise angles angles are then uniquely determined by the geometry of points in a plane.\footnote{The proof of this is an application of the Euler Characteristic formula:
\begin{equation}
V-E+F=2\,.
\end{equation}
The number of vertices $V$ is just the number of particles in the jet, $M$.  The number of faces $F$ is equal to the number of triangles that tesselate the plane, with vertices located at the particles.  This is $F=M-1$, as we include the face outside the region where the points are located.  It then follows that the number of edges $E$, that is, the number of pairwise angles necessary to uniquely specify their distribution, is $E=2M-3$.}  To determine all of these phase space variables, we extend the set of $N$-subjettinesses that were measured in the 2- and 3-body case.  In this case, the $3M-4$ observables we measure are:
\begin{equation}
\left\{
\tau_1^{(0.5)},\tau_1^{(1)},\tau_1^{(2)},\tau_2^{(0.5)},\tau_2^{(1)},\tau_2^{(2)},\dotsc,\tau_{M-2}^{(0.5)},\tau_{M-2}^{(1)},\tau_{M-2}^{(2)},\tau_{M-1}^{(1)},\tau_{M-1}^{(2)}
\right\}\,.
\end{equation}
Note that there are $3(M-2)+2=3M-4$ observables, and these will span the space of phase space variables for generic momenta configurations, when all particles have non-zero energy and are a finite angle from one another.

As we observed in the 3-body phase space case, for a collection of $M$ particles, all but one of the axes for the measurement of $(M-1)$-subjettiness lies along the direction of a particle.  Therefore, we only measure two $(M-1)$-subjettiness observables.  Stepping back another clustering as relevant for $(M-2)$-subjettiness, there are two possibilities:
\begin{itemize}
\item Either $M-3$ axes lie along the direction of $M-3$ particles in the jet, and the three remaining particles are all clustered around the last axis.  Then, the measurement of $(M-2)$-subjettiness is sensitive to the phase space configuration of 3 particles in the jet.  By measuring three $(M-2)$-subjettinesses and two $(M-1)$-subjettinesses, this then completely specifies the phase space configuration of those three particles.  

\item The other possibility is that $M-4$ axes lie along particles in the jet, while there are two particles clustered around each of the two remaining axes.  About each axis, you are sensitive to the phase space configuration of two particles, which corresponds to a total of 4 phase space variables.  Additionally, you are sensitive to the relative contribution of the two pairs of particles to the total $(M-2)$-subjettiness value.  This configuration therefore is described by 5 phase space variables, and can be completely specified by the measurement of three $(M-2)$-subjettinesses and two $(M-1)$-subjettinesses.
\end{itemize}

This argument can be continued at further stages in the declustering.  Each time an axis is removed, three new phase space variables are introduced.  These can be completely specified by the measurement of three additional $N$-subjettiness observables.  This then proves that the collection of $N$-subjettiness observables given above uniquely determines $M$-body phase space.

\end{itemize}

In the next section, we will study the information contained in this basis and use it to identify the features that are exploited in the discrimination of hadronically decaying $Z$ boson jets from QCD jets.

\section{Deep Learning Implementation}\label{sec:deeplearn}

In this section, we describe our event simulation and implementation of machine learning to the $N$-subjettiness basis of observables introduced in the previous section.  We generate $pp\to Z+$ jet and $pp\to ZZ$ events at the 13 TeV LHC with MadGraph5 v2.5.4 \cite{Alwall:2014hca}.  The $Z$ boson in $pp\to Z+$ jet events is decayed to neutrinos, while one $Z$ boson in $pp\to ZZ$ events is decayed to neutrinos, while the other is decayed to quarks.  These tree-level events are then showered in Pythia v8.223 \cite{Sjostrand:2006za,Sjostrand:2014zea} with default settings.  In \App{app:herwig}, we will show results showered with Herwig v7.0.4 \cite{Bahr:2008pv,Bellm:2015jjp}, however with one-tenth the number of events as the Pythia samples.  Ignoring the neutrinos in the showered and hadronized events, we use FastJet v3.2.1 \cite{Cacciari:2011ma,Cacciari:2005hq} to cluster the jets.  On the clustered anti-$k_T$ \cite{Cacciari:2008gp} jets with radius $R=0.8$ and minimum $p_T$ of 500 GeV, we then measure the basis of $N$-subjettiness observables using the code provided in FastJet contrib v1.026.  We emphasize that observables are measured on the particles as a proof of concept; we do not apply any detector simulation.

The precise set of observables we measure on the jet that we use for discrimination are the following.  We measure the jet mass and the collection of $N$-subjettiness observables sufficient to completely determine up through 6-body phase space.  That is, we measure the collection of $N$-subjettiness observables defined with $k_T$ axes:
\begin{equation}
\left\{
\tau_1^{(0.5)},\tau_1^{(1)},\tau_1^{(2)},\tau_2^{(0.5)},\tau_2^{(1)},\tau_2^{(2)},\tau_{3}^{(0.5)},\tau_{3}^{(1)},\tau_3^{(2)},\tau_4^{(0.5)},\tau_4^{(1)},\tau_4^{(2)},\tau_5^{(1)},\tau_5^{(2)}
\right\}\,.
\end{equation}
We will see that this collection of $N$-subjettiness observables is more than sufficient to describe all of the information useful for discrimination in the jet.  Additionally, for comparison, we will measure a collection of standard observables that have been defined for discrimination of boosted, hadronic decays of $Z$ bosons from jets initiated by QCD.  We measure  the $N$-subjettiness ratios $\tau_{2,1}^{(1)}$ and $\tau_{2,1}^{(2)}$ with one-pass winner-take-all (WTA) axes \cite{Bertolini:2013iqa,Larkoski:2014uqa,Larkoski:2014bia}, and (generalized) energy correlation function ratios $D_2^{(1)}$ and $D_2^{(2)}$ \cite{Larkoski:2014gra} and $N_2^{(1)}$ and $N_2^{(2)}$ \cite{Moult:2016cvt}.  The discrimination power of these observables will provide a benchmark for the information extracted in the machine learning of the collection of $N$-subjettiness observables.

All deep learning analysis was carried out on the NVIDIA DIGITS DevBox, with four GeForce GTX TitanX GPUs, built on the 28 nm Maxwell architecture. The specifications of the GPU are listed in Table~\ref{specs}. Only one GPU was used during training and testing. 
\begin{center}
	\begin{longtable}{cccccc}
		\toprule
		\thead{\shortstack{CUDA\\cores}} &
		\thead{\shortstack{Base/Boost.\\clock (MHz)}} & 
		\thead{\shortstack{Memory size\\(GB)}} & 
		\thead{\shortstack{Memory\\clock (Gbps)}}& 
		\thead{\shortstack{Interface\\width}}   & 
		\thead{\shortstack{Memory\\Bandwidth\\(GB/s)}}\\
		\midrule
		3072 & 1000/1075 & 12 & 7.0 & 384-bit & 336.5\\
		\bottomrule
		\caption{Manufacturer specifications of the GTX TitanX.\label{specs}}
	\end{longtable}
\end{center}
The dataset consisted of 7,868,000 events, split evenly between $Z$ and QCD jets, stored in the compressed HDF5 format  \cite{hdf5}.  The data was shuffled to ensure each data file had approximately a 1:1 ratio of both classes of events.  No mass cuts were imposed on the events fed to networks with the expectation that they would automatically learn the optimal cuts on mass and the observable phase space. The training and validation data consisted of 6,144,000 events and 1,536,000 events respectively, while 188,000 events were set aside for predictions.  All networks were trained using the highly modular Keras \cite{chollet2015keras} deep learning libraries and tested using the relevant scikit-learn \cite{scikit-learn} packages. At the time of training, data from the relevant columns of $N$-subjettiness variables was fed to the neural networks with the aid of a custom-designed data generator, which creates an archive of pre-processed data files. 

A single neural network architecture, consisted exclusively of five fully connected layers, was utilized for all analyses.  The first two Dense layers consisted of 10000 and 1000 nodes, respectively, and were assigned a Dropout \cite{dropout} regularization of 0.2, while next two Dense layers consisted of 100 nodes each, and were assigned a Dropout regularization of 0.1 to prevent over-fitting on training data by making each node more `independent'. The input layer and all hidden layers utilized the ReLU activation function \cite{conf/icml/NairH10}, while the output layer, consisting of a single node, used a sigmoid activation. The network was compiled with the binary cross-entropy loss minimization function, using the Adam optimization \cite{DBLP:journals/corr/KingmaB14}. Models were trained with Keras' default EarlyStopping, with a patience threshold of 5, to negate possible over-fitting.  For each set of observables, the typical number of training epochs was about 60.  To further eliminate errors due to under-training or over-training of networks, the same architecture was trained 25 different times for each round of analysis. The model that trained best for a given variable basis was picked based on a metric of maximizing the area under the signal vs.~background efficiency curve. 

\begin{figure}
\begin{center}
\includegraphics[width=.6\textwidth]{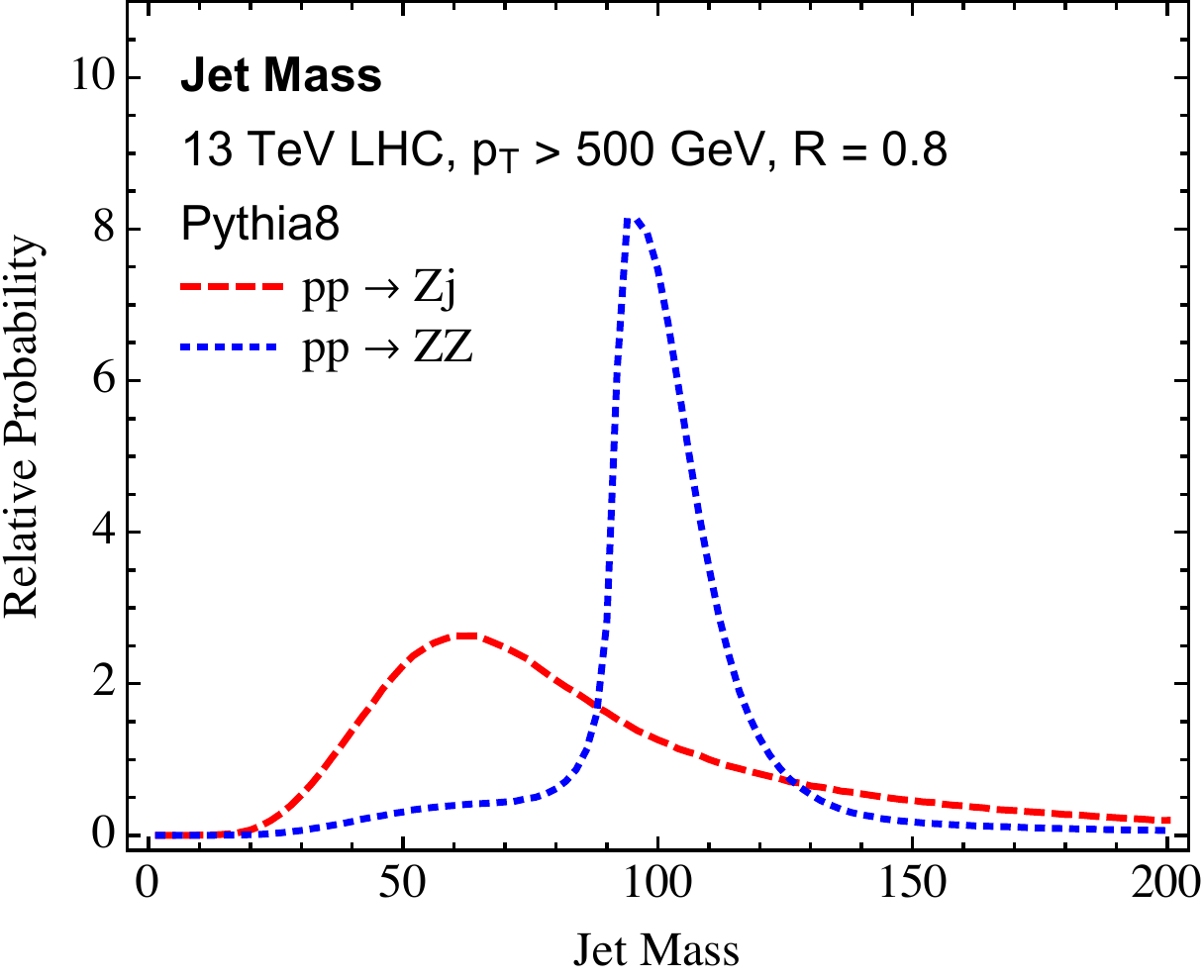}
\caption{
Distribution of the mass of the jet in $pp\to Zj$ (red dashed) and hadronically-decaying $Z$ boson in $pp\to ZZ$ (blue dotted) from the Pythia parton shower.  The minimum transverse momentum is 500 GeV, and the jets are found with the anti-$k_T$ algorithm with radius $R=0.8$.
}
\label{fig:massplot}
\end{center}
\end{figure}

\begin{figure}
\begin{center}
\subfloat[]{\label{fig:t21}
\includegraphics[width=7cm]{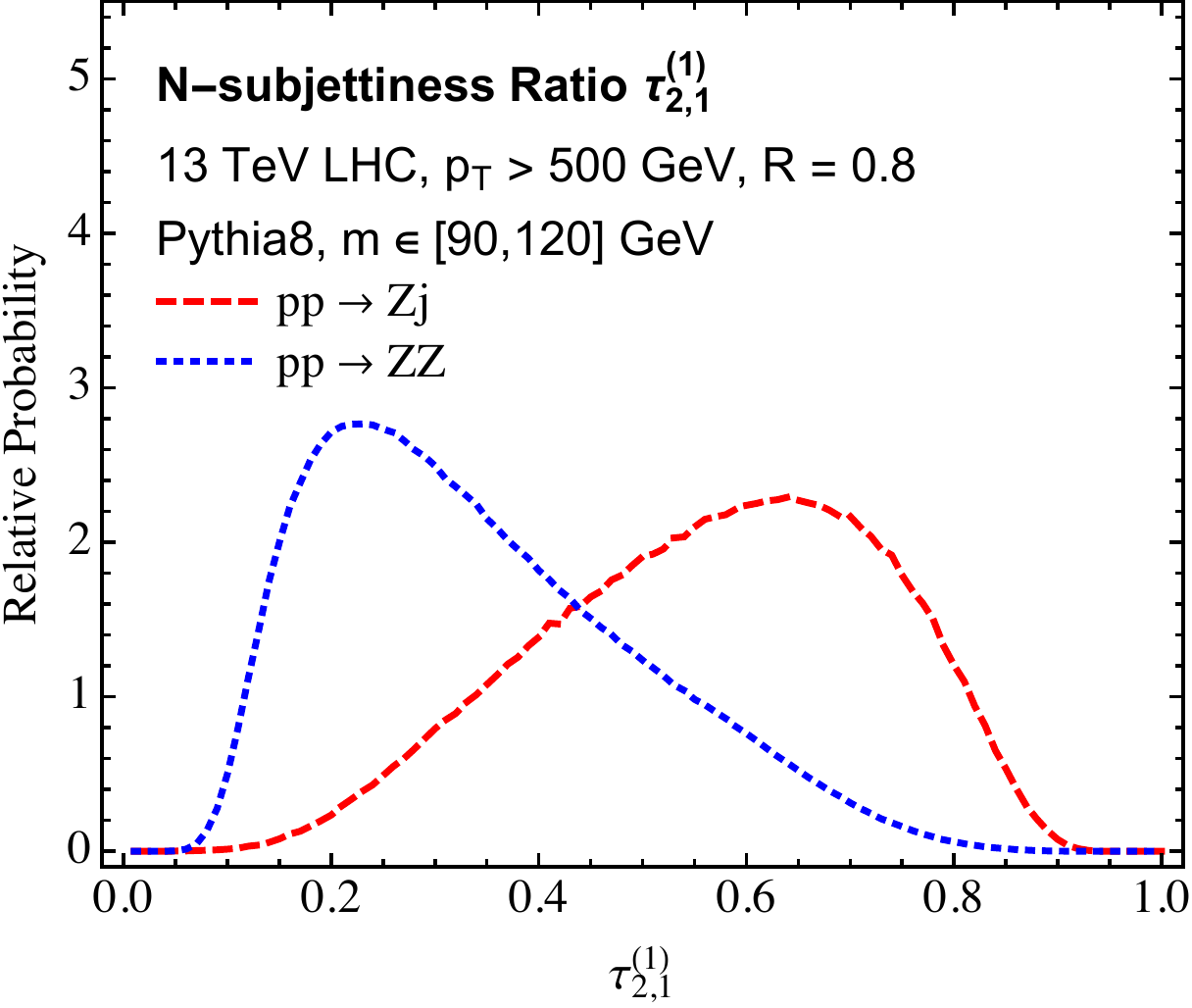}    
}\qquad
\subfloat[]{\label{fig:t22}
\includegraphics[width=7cm]{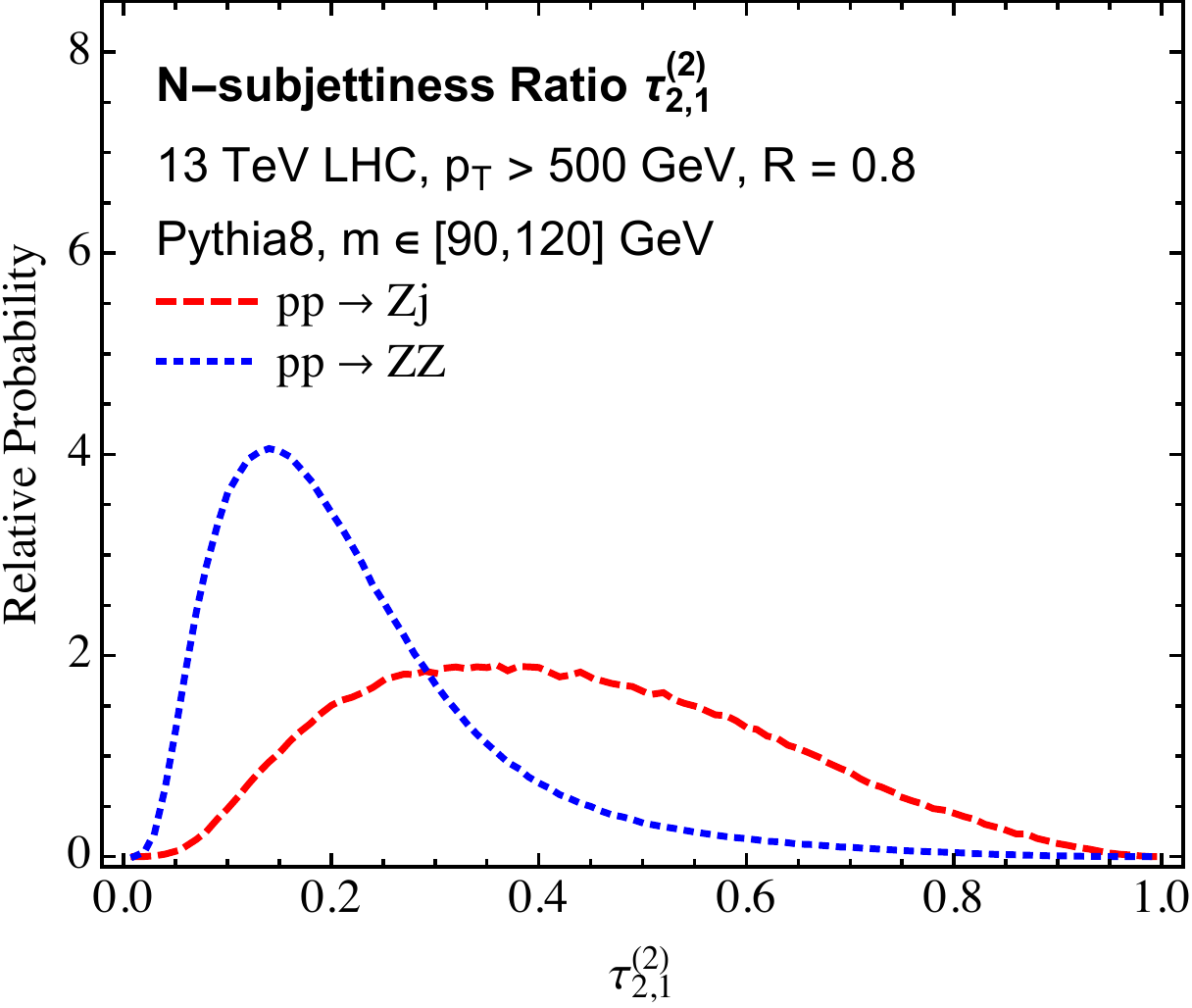}
}\\
\subfloat[]{\label{fig:d21}
\includegraphics[width=7cm]{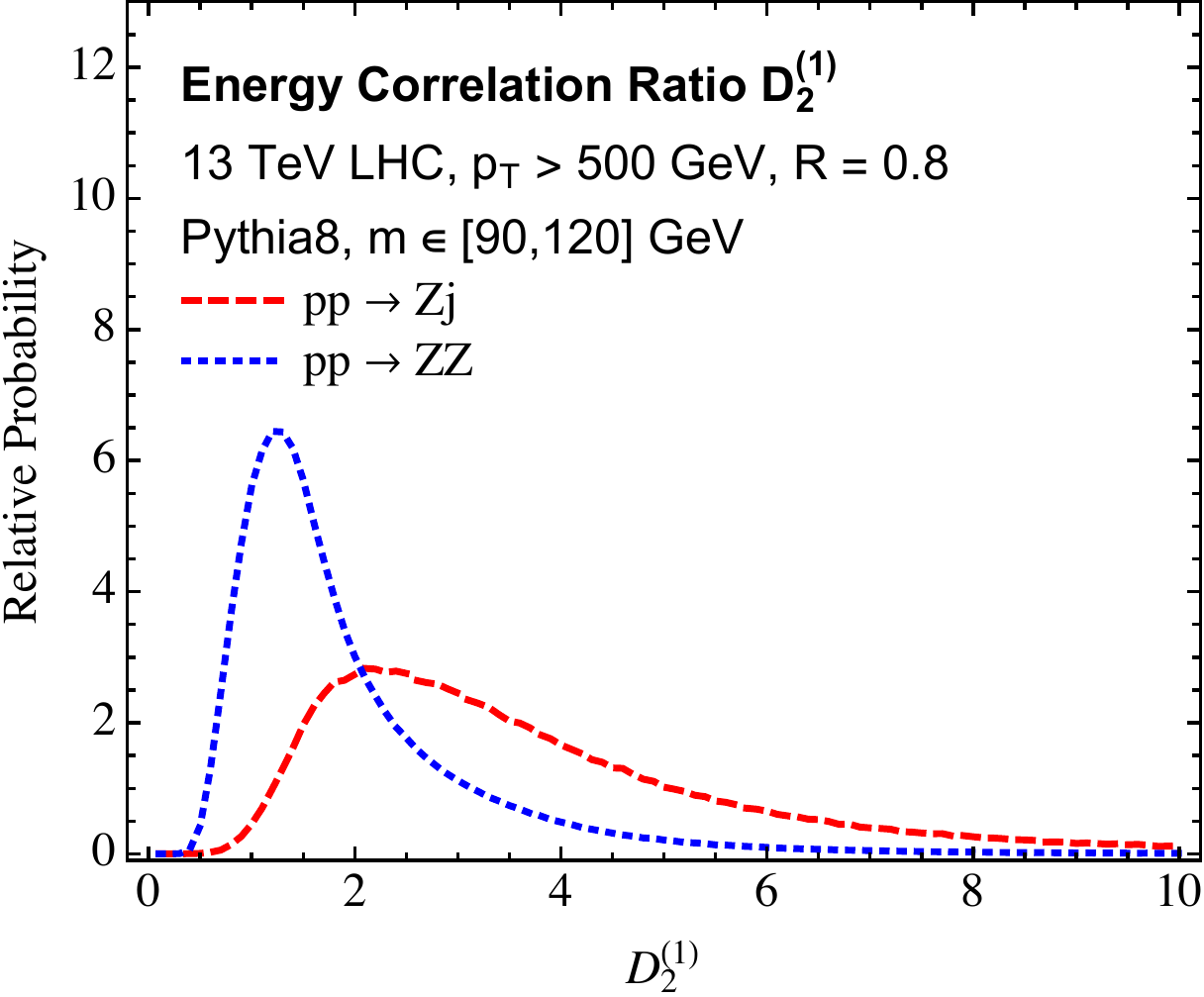}    
}\qquad
\subfloat[]{\label{fig:d22}
\includegraphics[width=7cm]{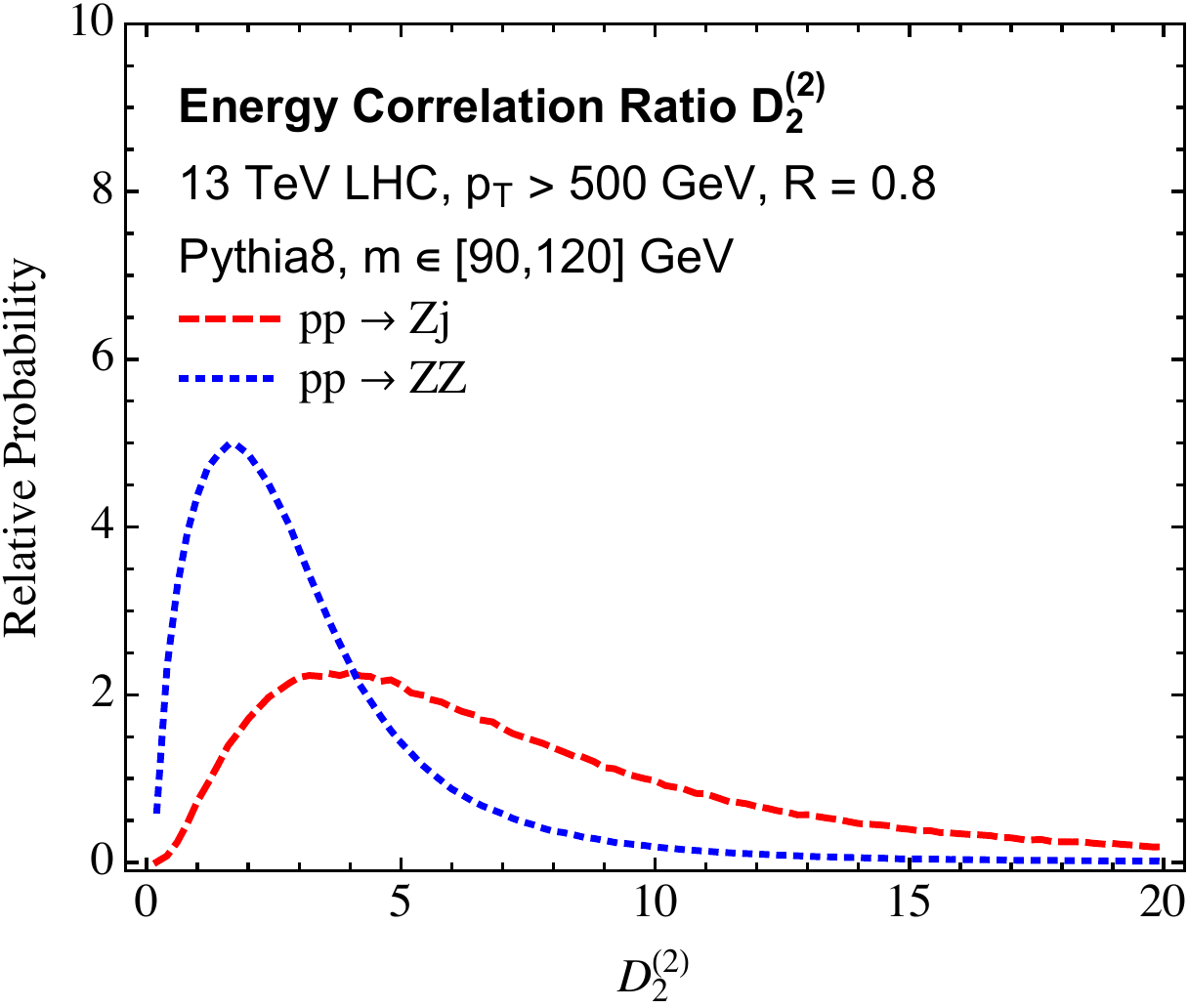}
}\\
\subfloat[]{\label{fig:n21}
\includegraphics[width=7cm]{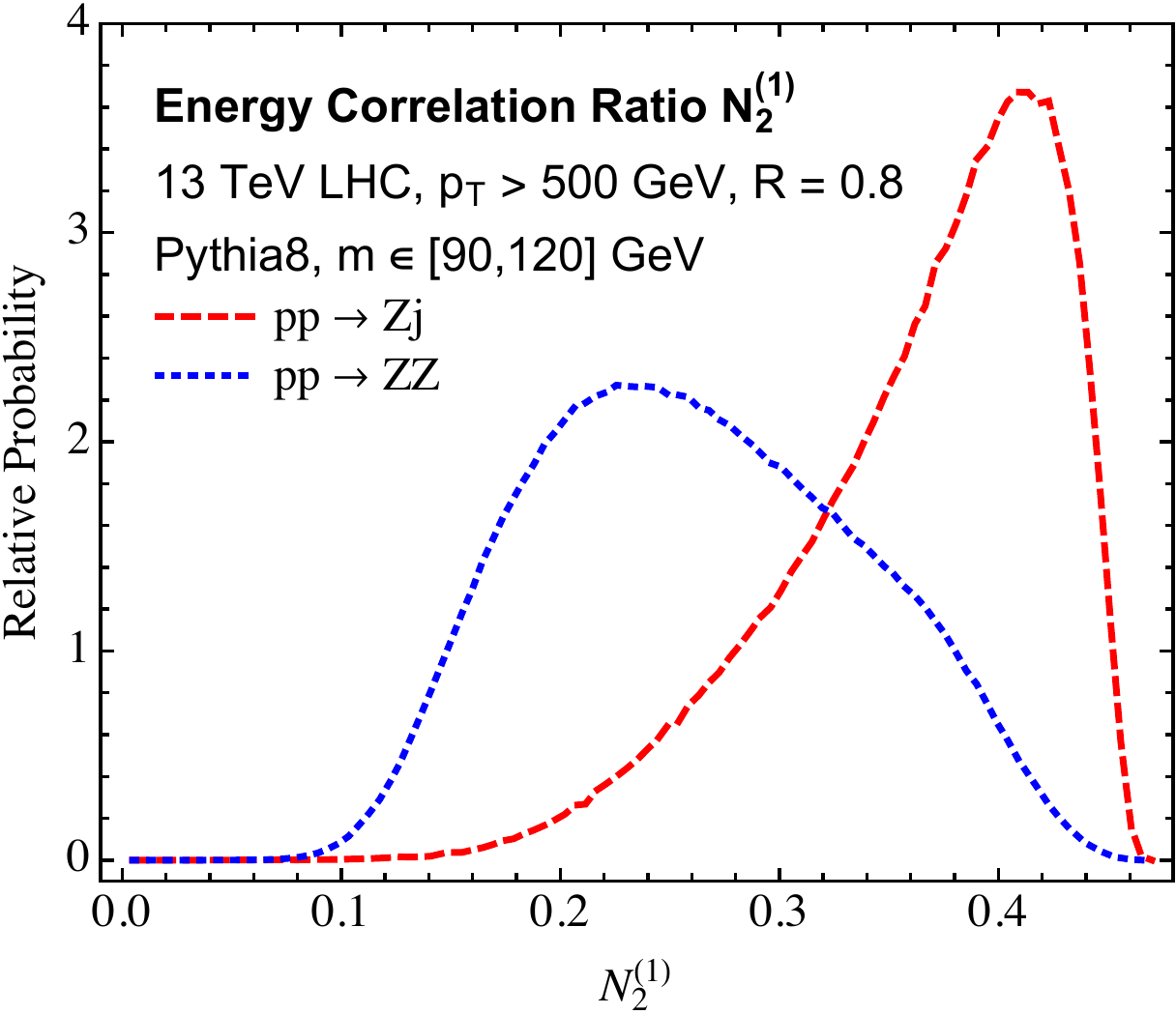}    
}\qquad
\subfloat[]{\label{fig:n22}
\includegraphics[width=7cm]{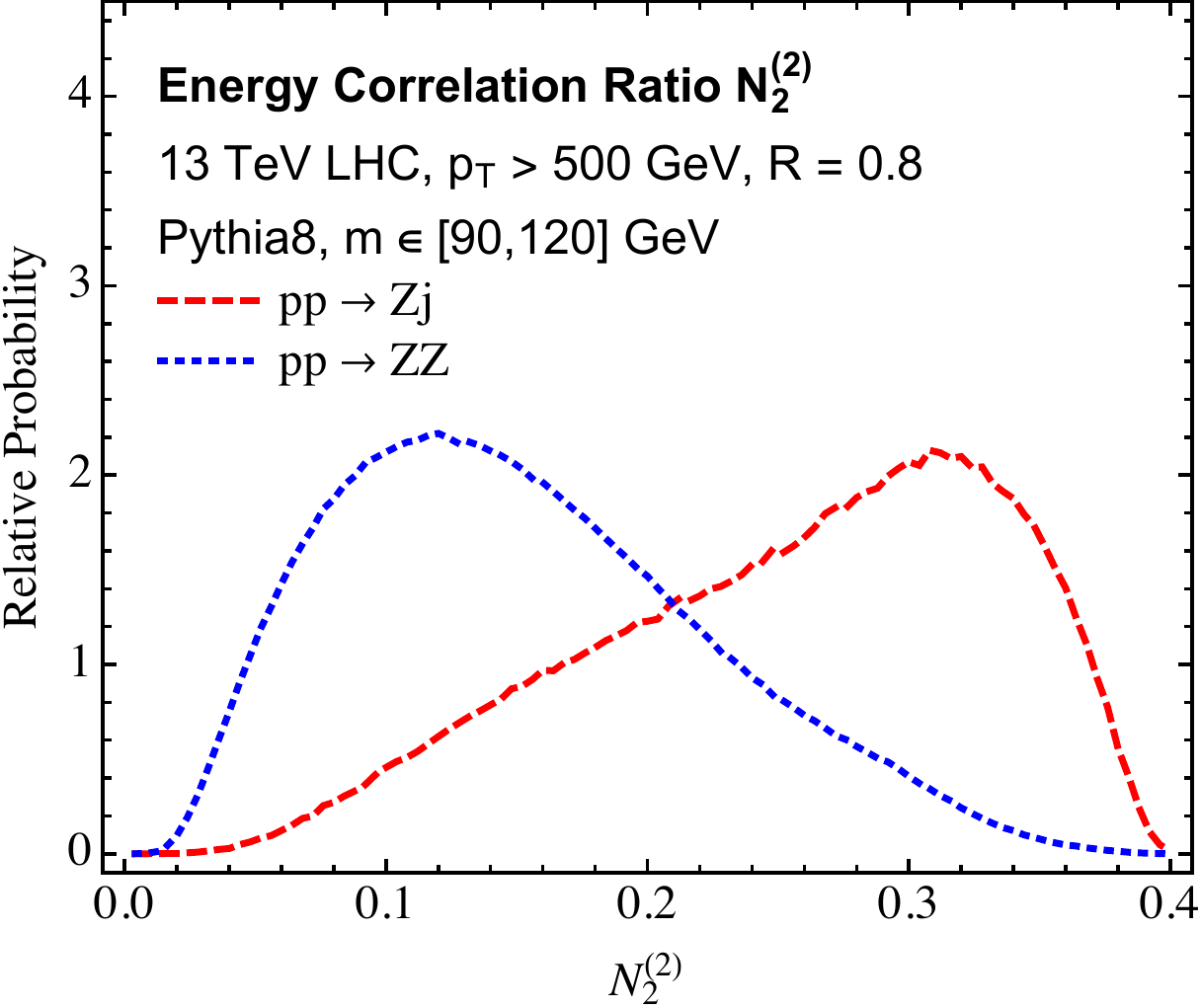}
}
\end{center}
\caption{
Distributions of various two-prong discrimination observables measured on the sample of jets showered with Pythia, on which a mass cut of $m\in[90,120]$ GeV has been placed.  From top to bottom are plotted signal (blue dotted) and background (red dashed) distributions of: $N$-subjettiness ratios $\tau_{2,1}^{(1)}$ (left) and $\tau_{2,1}^{(2)}$ (right), energy correlation function ratios $D_{2}^{(1)}$ (left) and $D_{2}^{(2)}$ (right), and $N_{2}^{(1)}$ (left) and $N_{2}^{(2)}$ (right).
}
\label{fig:obsdistros}
\end{figure}

Before showing the results from the deep neural network, we first show plots of the collection of observables sensitive to two-prong structure measured on the jets.  In \Fig{fig:massplot}, we plot the mass of the signal and background jets as defined by the simulation and jet finding from earlier.  Applying a mass cut around the $Z$ boson peak, we then measure the two-prong jet observables.  In \Fig{fig:obsdistros}, we show the distributions of the $N$-subjettiness and energy correlation function ratios $\tau_{2,1}^{(\beta)}$, $D_2^{(\beta)}$, and $N_2^{(\beta)}$.  As was extensively studied in the original works, these plots make clear the separation power that these observables enable.  When we compare these observables to the discrimination power of the $M$-body phase space observables, we relax the hard mass cut, and let the machine learn the optimal mass and observable cuts dynamically.

In \Fig{fig:introplot}, we plot the signal jet ($Z$ boson) efficiency versus the background jet (QCD)  rejection rate for the collection of observables that minimally span $M$-body phase space, along with the jet mass.  The observables that are passed to the neural network to specify $M$-body phase space are, explicitly:
\begin{align*}
&\text{2-body: } \tau_1^{(1)}\,,\tau_1^{(2)}\\
&\text{3-body: } \tau_1^{(0.5)}\,,\tau_1^{(1)}\,,\tau_1^{(2)}\,,\tau_2^{(1)}\,,\tau_2^{(2)}\\
&\text{4-body: } \tau_1^{(0.5)}\,,\tau_1^{(1)}\,,\tau_1^{(2)}\,,\tau_2^{(0.5)}\,,\tau_2^{(1)}\,,\tau_2^{(2)}\,,\tau_3^{(1)}\,,\tau_3^{(2)}\\
&\text{5-body: } \tau_1^{(0.5)}\,,\tau_1^{(1)}\,,\tau_1^{(2)}\,,\tau_2^{(0.5)}\,,\tau_2^{(1)}\,,\tau_2^{(2)}\,,\tau_3^{(0.5)}\,,\tau_3^{(1)}\,,\tau_3^{(2)}\,,\tau_4^{(1)}\,,\tau_4^{(2)}\\
&\text{6-body: } \tau_1^{(0.5)}\,,\tau_1^{(1)}\,,\tau_1^{(2)}\,,\tau_2^{(0.5)}\,,\tau_2^{(1)}\,,\tau_2^{(2)}\,,\tau_3^{(0.5)}\,,\tau_3^{(1)}\,,\tau_3^{(2)}\,,\tau_4^{(0.5)}\,,\tau_4^{(1)}\,,\tau_4^{(2)}\,,\tau_5^{(1)}\,,\tau_5^{(2)}
\end{align*}
Significant gains in discrimination power are observed by including observables sensitive to higher-body phase space, until enough observables to specify at least 4-body phase space are included.  Including observables sensitive to 5- and 6-body phase space does not improve discrimination power, and therefore suggests that there is only an extremely limited amount of information in a jet useful for discrimination.

\begin{figure}
\begin{center}
\subfloat[]{\label{fig:t21roc}
\includegraphics[width=7cm]{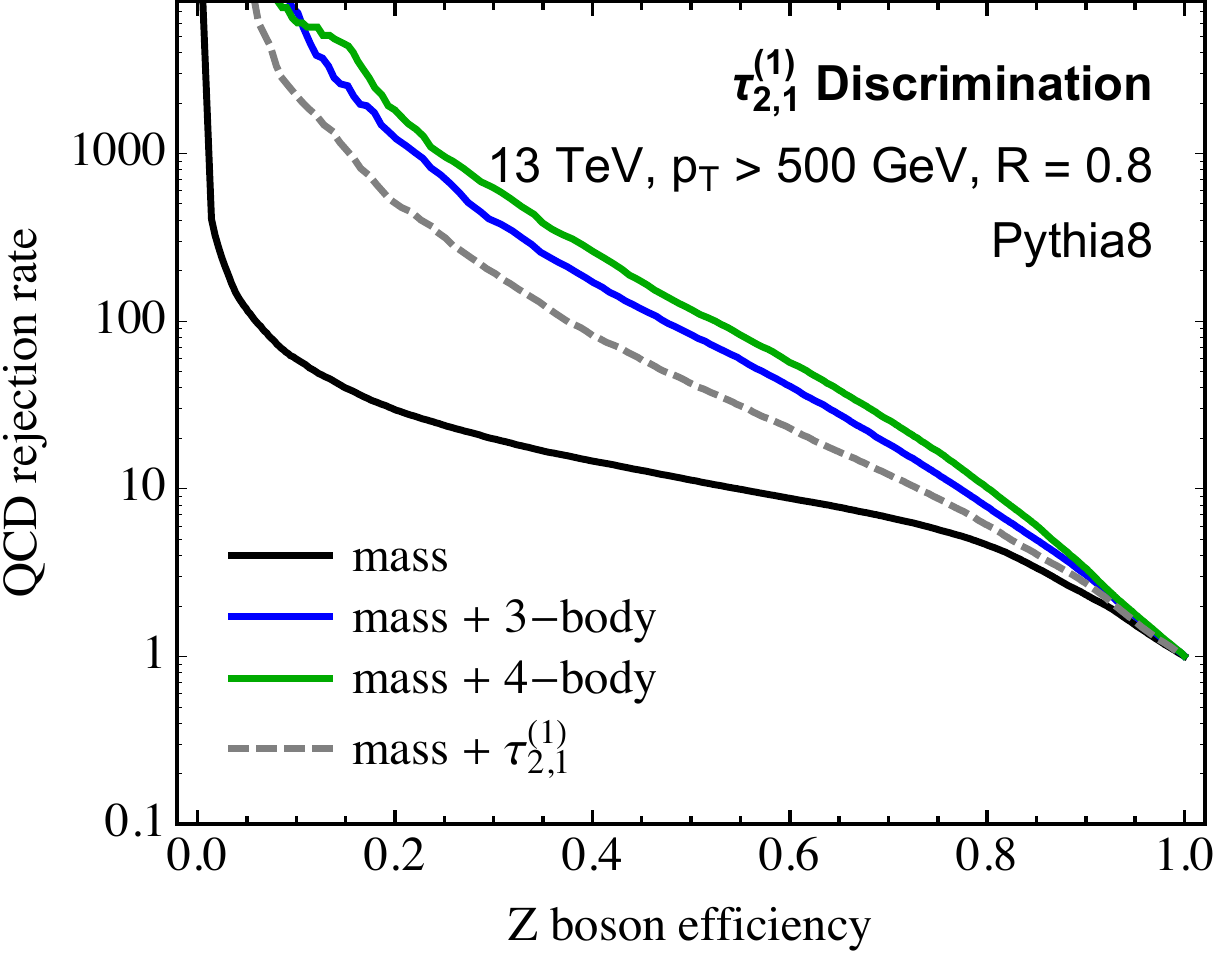}    
}\qquad
\subfloat[]{\label{fig:t22roc}
\includegraphics[width=7cm]{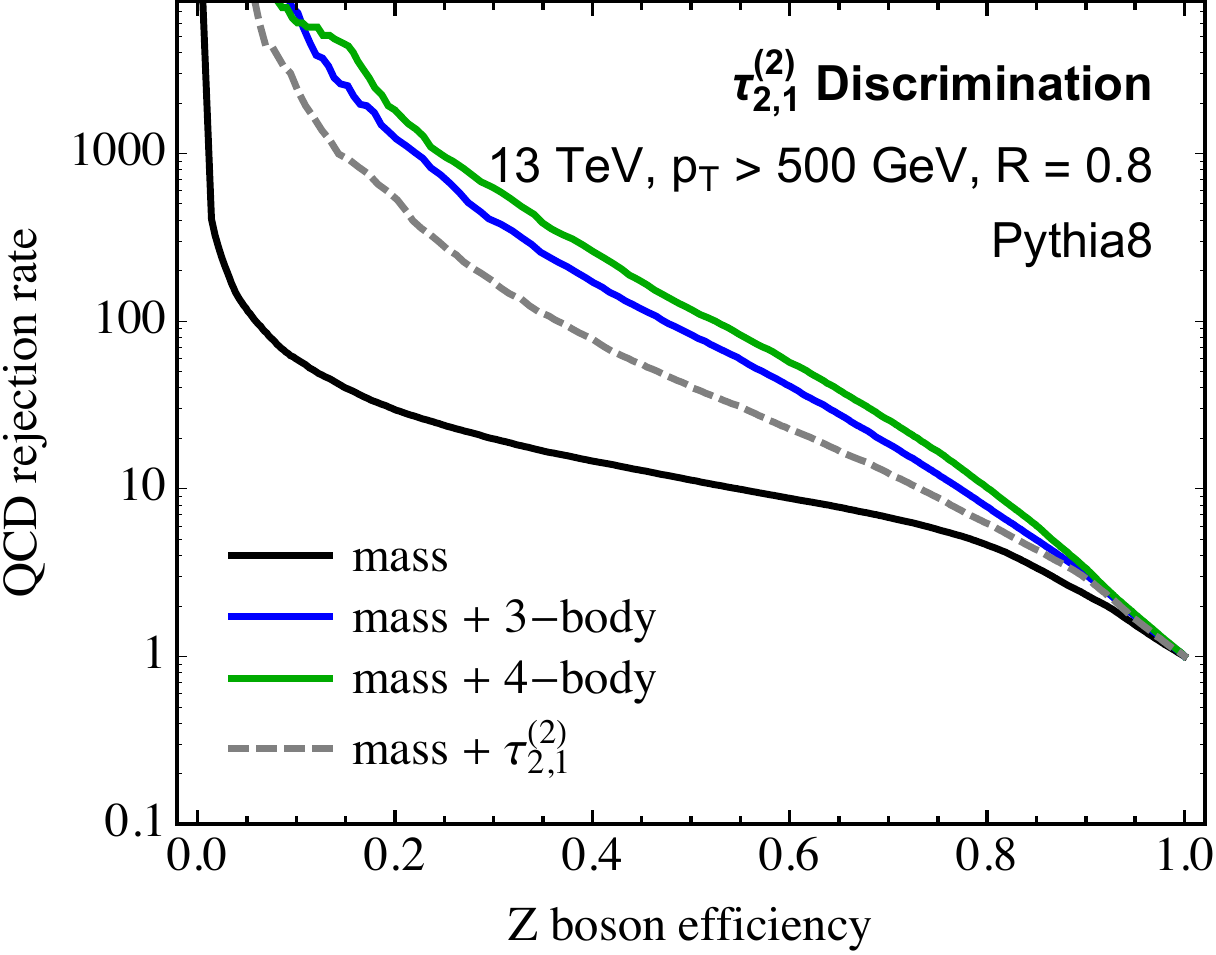}
}\\
\subfloat[]{\label{fig:d21roc}
\includegraphics[width=7cm]{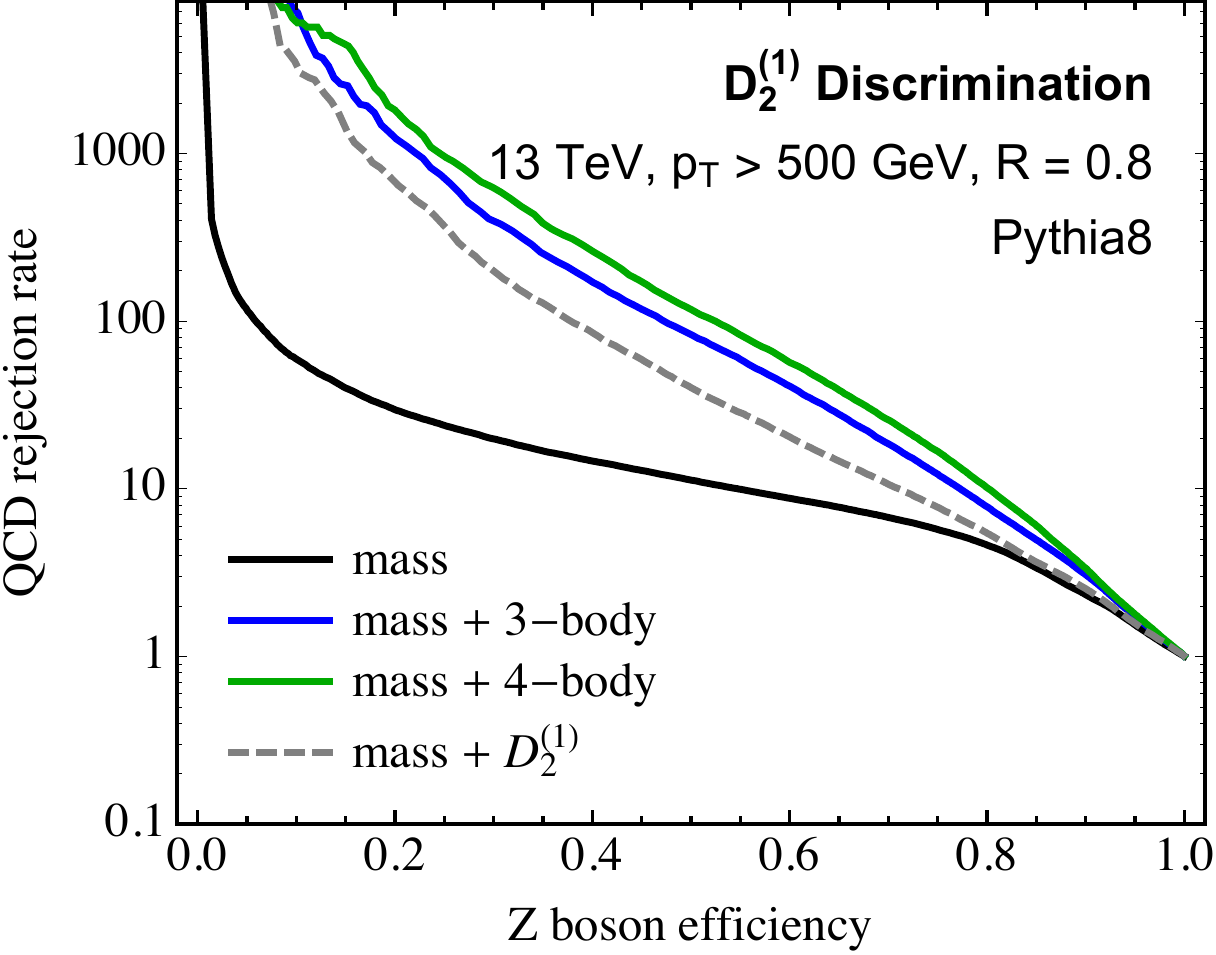}    
}\qquad
\subfloat[]{\label{fig:d22roc}
\includegraphics[width=7cm]{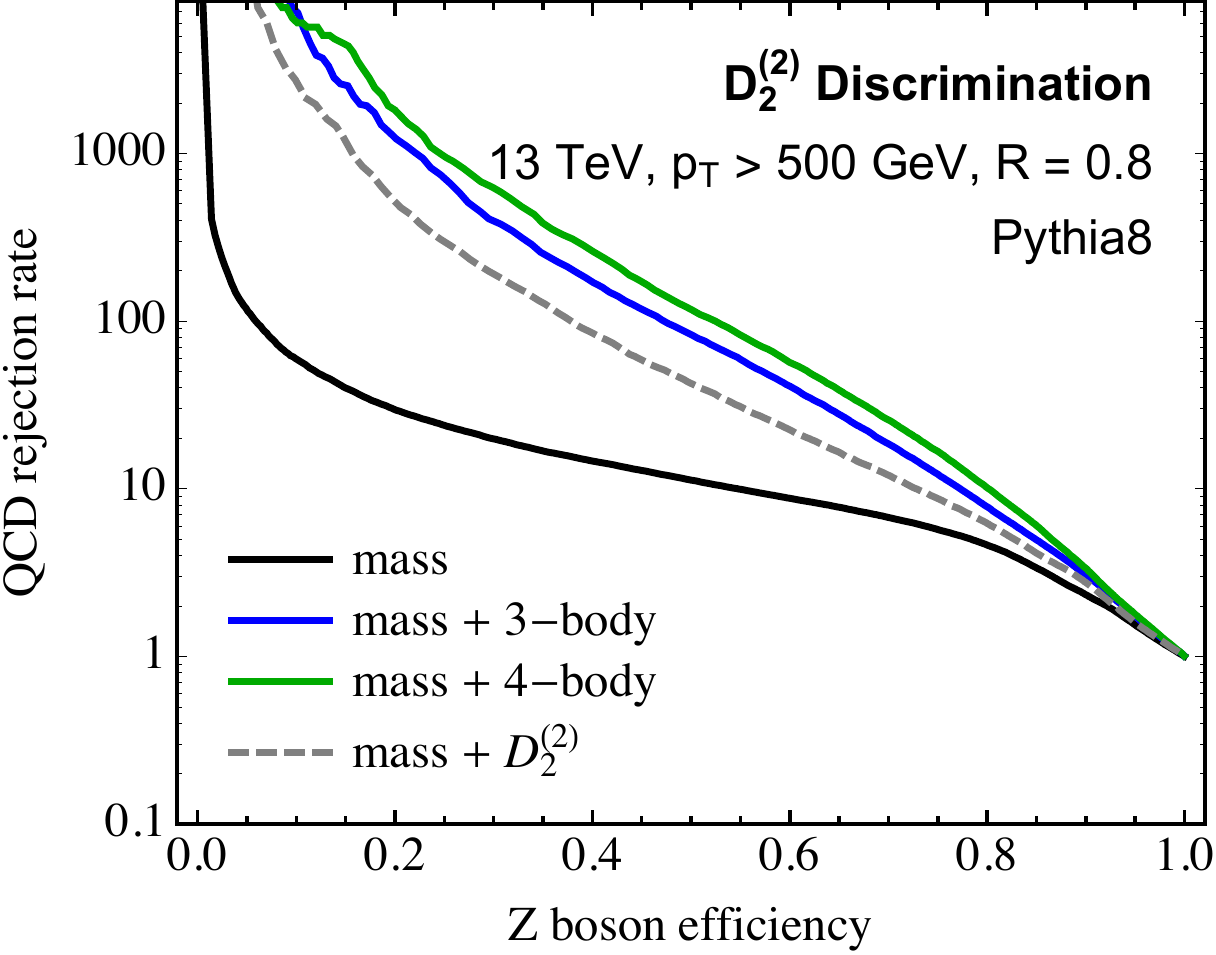}
}\\
\subfloat[]{\label{fig:n21roc}
\includegraphics[width=7cm]{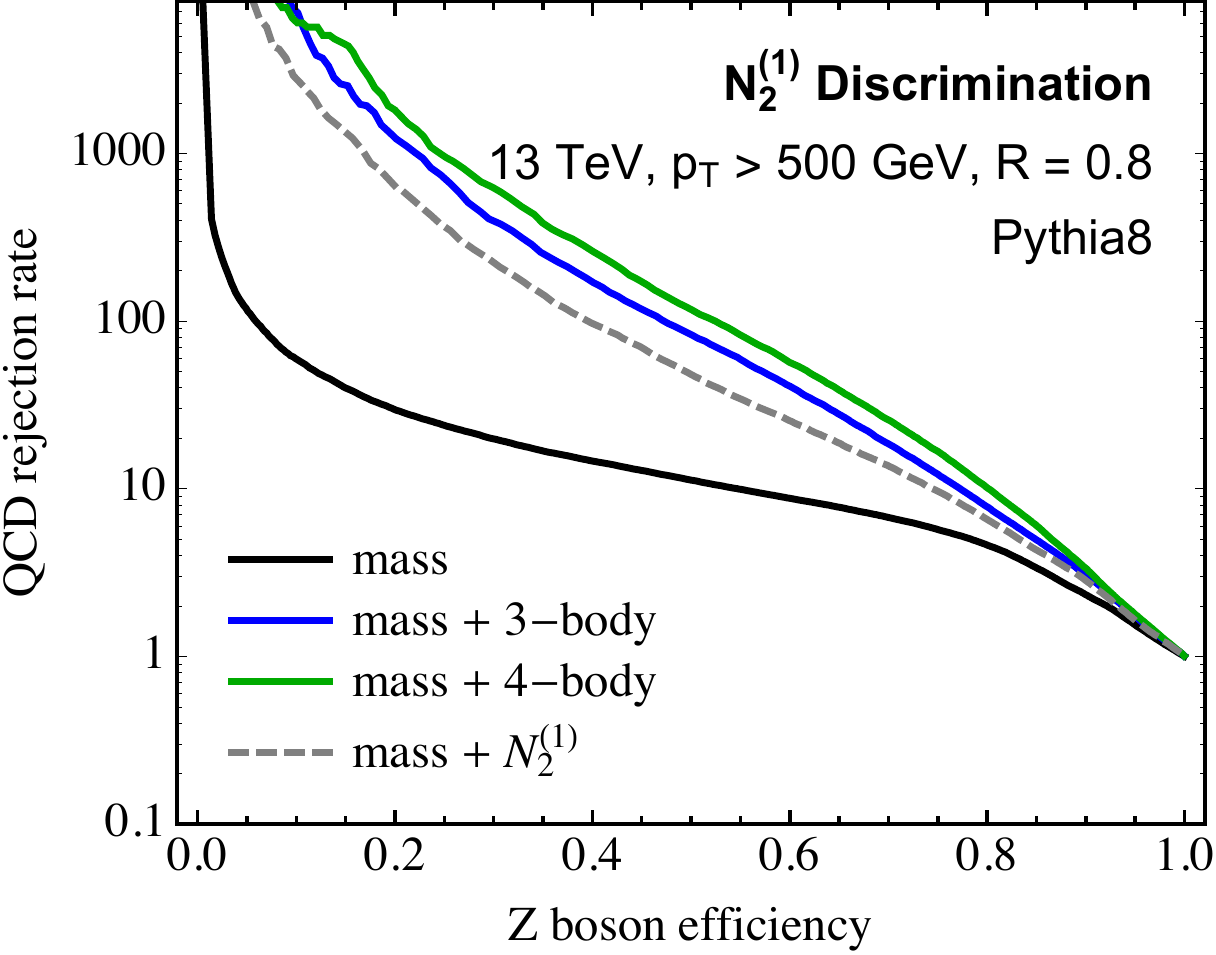}    
}\qquad
\subfloat[]{\label{fig:n22roc}
\includegraphics[width=7cm]{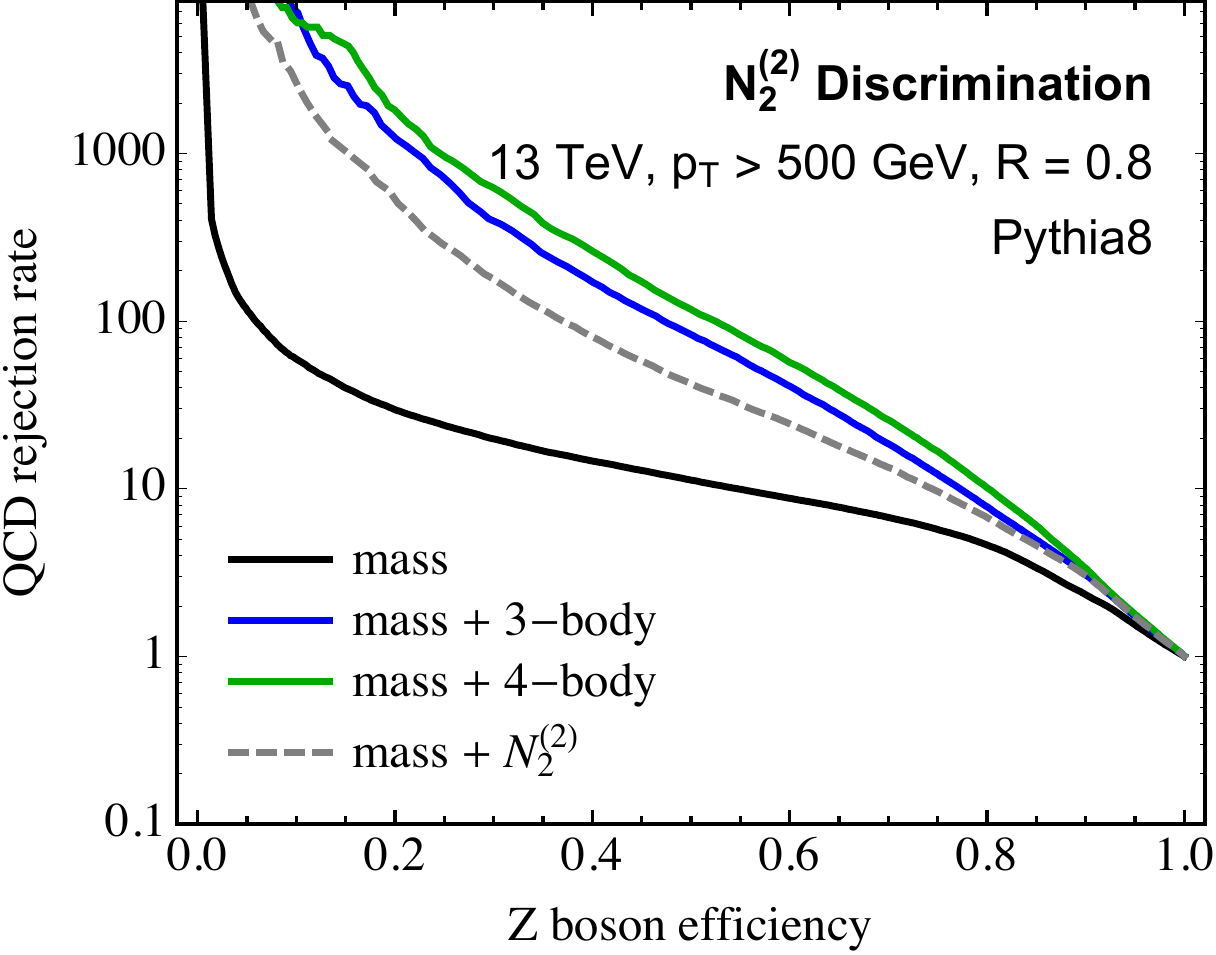}
}
\end{center}
\caption{
Signal efficiency versus background rejection rate for jet mass plus $N$-subjettiness ratios $\tau_{2,1}^{(1)}$ (left) and $\tau_{2,1}^{(2)}$ (right), energy correlation function ratios $D_{2}^{(1)}$ (left) and $D_{2}^{(2)}$ (right), and $N_{2}^{(1)}$ (left) and $N_{2}^{(2)}$ (right), as determined by the neural network.  For comparison, we also include the signal efficiency versus background rejection rate for jet mass, jet mass plus 3-body phase space observables, and jet mass plus 4-body phase space observables.
}
\label{fig:obsroc}
\end{figure}

To see what information is necessary to accomplish the maximal discrimination power, in \Fig{fig:obsroc} we plot the signal efficiency versus background rejection rate for the collection of $N$-subjettiness and energy correlation function ratios plotted earlier.  For comparison, we also include the corresponding curves for the jet mass, jet mass plus 3-body phase space observables, and jet mass plus 4-body phase space observables.  The discrimination power of all of these observables are comparable, and this illustrates that they appear to capture most of the information contained in the 3-body phase space observables.  Then, to match the maximum discrimination power (as represented by the jet mass plus 4-body phase space curve), one just needs to augment the measurement of jet mass and an $N$-subjettiness or energy correlation function ratio with observables that are sensitive to some 3- and 4-body phase space information.  We leave the construction of these optimal 3- and 4-body phase space observables for this purpose to future work.

\begin{figure}
\begin{center}
\includegraphics[width=.6\textwidth]{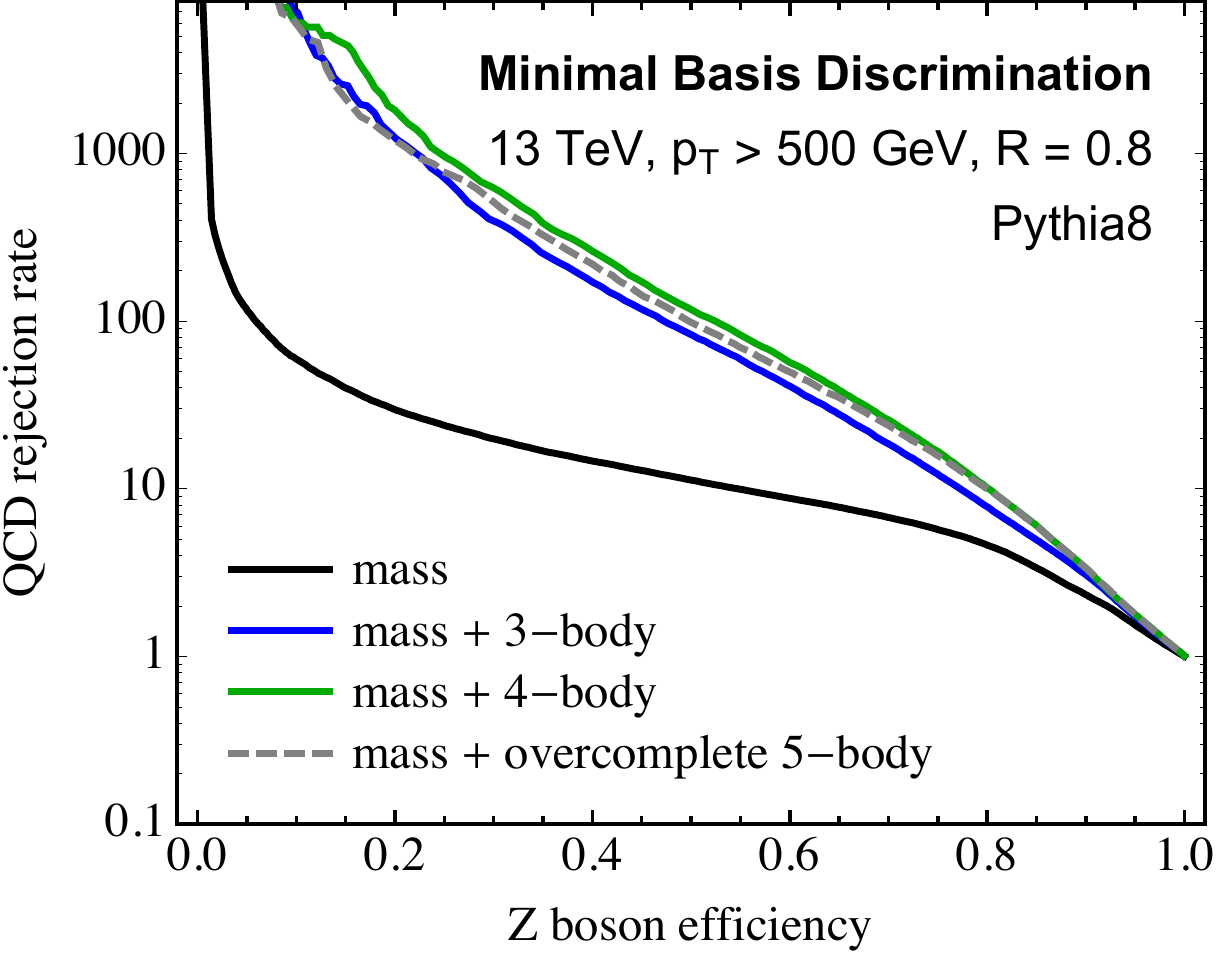}
\caption{
Signal efficiency versus background rejection rate for jet mass plus the overcomplete basis of observables that are sensitive to 5-body phase space described in the text, as determined by the neural network.  For comparison, we also include the signal efficiency versus background rejection rate for jet mass, jet mass plus minimal 3-body phase space observables, and jet mass plus the minimal 4-body phase space observables.
}
\label{fig:overplot}
\end{center}
\end{figure}

As a cross check that our minimal basis of $N$-subjettiness observables listed above does capture the maximal amount of information useful for discrimination, in \Fig{fig:overplot}, we compare our minimal basis to an overcomplete basis of observables.  Here, we measure the mass and the following collection of $N$-subjettiness observables on the jet:
\begin{align}
&\left\{
\tau_1^{(0.25)},\tau_1^{(0.5)},\tau_1^{(1)},\tau_1^{(2)},\tau_1^{(4)},\tau_2^{(0.25)},\tau_2^{(0.5)},\tau_2^{(1)},\tau_2^{(2)},\tau_2^{(4)},\tau_3^{(0.25)},\tau_{3}^{(0.5)},\tau_{3}^{(1)},\tau_3^{(2)},\tau_3^{(4)},\right.\\
&
\hspace{10cm}
\left.\tau_4^{(0.25)},\tau_4^{(0.5)},\tau_4^{(1)},\tau_4^{(2)},\tau_4^{(4)}
\right\}\,. \nonumber
\end{align}
From our arguments in \Sec{sec:basis}, this is an overcomplete basis for 5-body phase space and therefore should not contain any additional information useful for discrimination.  This is illustrated in \Fig{fig:overplot} where we plot the discrimination power of this overcomplete basis as determined by the neural network described earlier.  For comparison, we also show the discrimination power of the jet mass, the jet mass plus the 3-body observable basis, and the jet mass plus the 4-body observable basis as determined by the neural network described earlier.  As expected, no improvement of discrimination power is accomplished when more observables beyond the minimal set are included.  The apparent slight decrease in discrimination power using the overcomplete basis is likely due to suboptimal training because of the large number of input observables.

In \App{app:morearchs}, we present results for the signal vs.~background efficiency as determined by a neural network with an additional hidden layer and the result of a boosted decision tree.  These different classification networks demonstrate the same conclusion, that discrimination power saturates once enough observables are measured to resolve 4-body phase space.  Additionally, these results show that the discrimination power of the overcomplete basis is just marginally better than that accomplished by the 4-body observable basis.  This is consistent with our observation that 4-body phase space is essentially saturating all useful discrimination information.

\section{Conclusions}\label{sec:conc}

Motivated by both the enormous data sets produced by the ATLAS and CMS experiments as well as their exceptional resolution, deep learning approaches to physics at the LHC are seeing an increased interest.  This is especially true for jet physics, where the identification of the initiating particle of a jet is of fundamental importance.  Previous applications of deep learning to jet physics applied techniques from computer science (like image recognition or natural language processing) and demonstrated impressive discrimination power.  While the effectiveness of these methods is exceptional, they often lack a physical interpretation and are not presented in a constructive manner.  The deep neural network is definitely identifying relevant structure in the jets, but what this is or if it is just a feature of the simulated data is not identified.  Other recent efforts to reduce dependence on modeling have been studied in the context of weak supervision in Ref.~\cite{Dery:2017fap}.

In this paper, we have approached the problem of machine learning for jet physics in a physically clear, constructive manner.  Instead of providing the machine with the energy deposits in calorimeter cells of the jet, we measure a basis of observables on the jet that completely and minimally spans $M$-body phase space.  The effective resolution to the emissions in the jet is increased by increasing the number of observables measured on the jet.  We demonstrated that the information useful for discrimination of a jet initiated by a boosted, hadronically-decaying $Z$ boson from a jet initiated by a light QCD parton is saturated when enough observables are measured to span 4-body phase space.  As 4-body phase space is only 8 dimensional, the amount of useful information in the jet is quite small.    Additionally, this procedure is constructive in the sense that one can then form observables that are non-zero for a jet with four constituents to optimally discriminate signal from background.  Similar constructions of observable bases for identifying particular phase space regions has been studied recently to resum non-global logarithms \cite{Larkoski:2015zka} and calculate multi-differential cross sections on jets \cite{Larkoski:2014tva,Procura:2014cba}.

Important for our analysis is that we use an IRC safe basis of observables that span the $M$-body phase space, namely, the $N$-subjettiness observables.  This is vital for constructibility, as in principle the cross section for the measurement of multiple $N$-subjettiness observables on a jet can be calculated in the perturbation theory of QCD.\footnote{Actually calculating distributions of $N$-subjettiness observables in practice may be a significant challenge, however \cite{Larkoski:2015uaa}.}  It would be possible to additionally include information that is not IRC safe, for example, jet charge.  Nevertheless, some non-IRC safe information is already included in this approach, like the jet constituent multiplicity.  Additionally, included in the basis of $M$-body phase space observables are techniques like jet grooming that systematically remove radiation from the jet.  This could enable a systematic study of how jet grooming methods affect the optimal discrimination observables, which has been addressed recently \cite{Moult:2016cvt,Salam:2016yht}.

An advantage of our approach is that the jet data is preprocessed in a useful way at the same time that the basis observables are being measured.  In applications of image processing to jets, one typically has to perform a series of transformations to ensure that different jets can be compared (see the discussion in, e.g., Ref.~\cite{deOliveira:2015xxd}).  Jets must be rotated and rescaled appropriately so that (approximate) symmetries do not wash out the ability to discriminate.  By instead measuring a collection of IRC safe observables like $N$-subjettiness on which we train, this preprocessing step is unnecessary, as the value of the observable is only sensitive to relative angles between particles and energy fractions.

From our results, it would also be interesting to study in detail the information for discrimination that is missed when using standard jet observables like $N$-subjettiness ratios $\tau_{2,1}^{(\beta)}$ or energy correlation function ratios $D_2^{(\beta)}$ or $N_2^{(\beta)}$.  The construction and justification of these particular observables exploited properties of QCD in the soft and/or collinear limits.  These observables appear to be sensitive to most of the 3-body phase space information available for discrimination of boosted, hadronically decaying $Z$ bosons from QCD jets.  Observables that are sensitive to the remaining information for discrimination could be constructed by studying in detail the differences between how the decays of $Z$ bosons and QCD fill 4-body phase space.  We anticipate that these methods can also be used for discrimination of many different types of jets, including quark versus gluon and QCD versus top quark discrimination, as well as for multi-label classification of jets.  The ultimate goal of such a program would be to design an anti-QCD tagger which could identify, using only a few observables that are sensitive to a small phase space, if a jet was likely initiated by a light QCD parton.  This could open the door to new classes of observables that are sensitive to exotic configurations within jets.

\acknowledgments

We thank Kyle Cranmer, Michael Kagan, Ian Moult, Ben Nachman, Duff Neill, Justin Pilot, Francesco Rubbo, Ariel Schwartzman, Jesse Thaler, and Daniel Whiteson for comments on the draft.  We also thank our anonymous referee for suggesting the studies presented in \App{app:morearchs}.

\appendix

\section{Explicit Expressions for 3-Body Phase Space}\label{app:3body}

In this appendix, we present the explicit expressions for the 1- and 2-subjettiness observables measured on a jet with three particles.  The configuration of particles in the jet is shown in \Fig{fig:3body}.  We will start with the evaluation of the 2-subjettiness observables, and then the 1-subjettiness observables.

\subsection{2-subjettiness}

For measuring 2-subjettiness, we identify two axes defined by the exclusive $k_T$ algorithm with $E$-scheme recombination.  For three particles, one of the axes must necessarily lie along the direction of one particle in the jet, which we can take to be particle 3 without loss of generality.  Then, only particles 1 and 2 can contribute to 2-jettiness.  Call the axis about which particles 1 and 2 are clustered $\hat A$.  Then, from \Fig{fig:3body}, the angle that particles 1 and 2 make with $\hat A$ are:
\begin{align}
&\theta_{1\hat A} = \frac{z_2}{z_1+z_2}\theta_{12}\,, & \theta_{2\hat A} = \frac{z_1}{z_1+z_2}\theta_{12}\,.
\end{align}
The 2-subjettiness observables that we measure are then:
\begin{align}
\tau_2^{(1)} &= z_1\cdot \frac{z_2}{z_1+z_2}\theta_{12} + z_2\cdot \frac{z_1}{z_1+z_2}\theta_{12} = \frac{2z_1 z_2}{z_1+z_2}\theta_{12}\,,\\
\tau_2^{(2)} &= z_1\left(\frac{z_2}{z_1+z_2}\theta_{12}\right)^2 + z_2\cdot \left(\frac{z_1}{z_1+z_2}\theta_{12}\right)^2=\frac{z_1z_2}{z_1+z_2}\theta_{12}^2\,.
\end{align}
Therefore the values of the 2-subjettiness observables can be inverted to determine the relative momentum fraction
$$
\frac{z_1z_2}{z_1+z_2}\,,
$$
and the pairwise angle $\theta_{12}$.

\subsection{1-subjettiness}

\begin{figure}
\begin{center}
 \includegraphics[width=.35\textwidth]{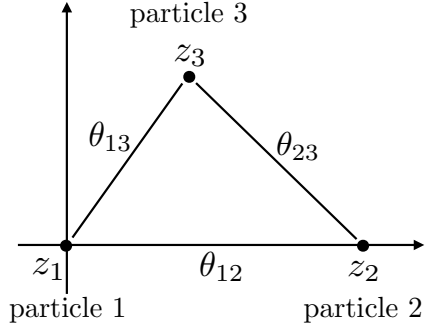}
\caption{
Configuration in a plane of a jet with three particles.  Pairwise angles $\theta_{ij}$ and momentum fractions $z_i$ of the individual particles are labeled.  Momentum conservation enforces that $z_3 = 1-z_1-z_2$.
}
\label{fig:3bodyplane}
\end{center}
\end{figure}

Now, we would like to calculate the value of 1-subjettiness on this configuration of particles.  This requires determining the angle between each of the three particles and their direction of net momentum.  To determine these angles, we consider the distribution of particles in the jet in a plane, as displayed in \Fig{fig:3bodyplane}.  We set particle 1 at the origin $(0,0)$ of the plane, particle 2 along the horizontal axis at $(\theta_{12},0)$, and particle 3 at a generic point in the plane.  The horizontal and vertical coordinates of particle 3 can be calculated to be:
\begin{equation}
\text{particle 3: } \left(
\frac{\theta_{12}^2+\theta_{13}^2-\theta_{23}^3}{2\theta_{12}},\frac{\sqrt{2\theta_{12}^2\theta_{23}^2+2\theta_{12}^2\theta_{13}^2+2\theta_{13}^2\theta_{23}^2-\theta_{12}^4-\theta_{23}^4-\theta_{13}^4}}{2\theta_{12}}
\right)\,.
\end{equation}

With this expression, we can determine the location of the jet axis.  With $E$-scheme recombination, the jet axis is located at the momentum-weighted centroid of the three particles:
\begin{equation}
\text{jet center: }\left(
z_2\theta_{12}+z_3 \frac{\theta_{12}^2+\theta_{13}^2-\theta_{23}^3}{2\theta_{12}}, z_3 \frac{\sqrt{2\theta_{12}^2\theta_{23}^2+2\theta_{12}^2\theta_{13}^2+2\theta_{13}^2\theta_{23}^2-\theta_{12}^4-\theta_{23}^4-\theta_{13}^4}}{2\theta_{12}}
\right)\,.
\end{equation}
Here, for conciseness, we express $z_3 = 1-z_1-z_2$.  It then follows that the angle from each particle to this jet axis $\hat A$ is:
\begin{align}
\theta_{1\hat A}^2 &= z_2^2\theta_{12}^2 + z_3^2\theta_{13}^2 +z_2 z_3 (\theta_{12}^2+\theta_{13}^2-\theta_{23}^2)\,,\\
\theta_{2\hat A}^2 &= z_1^2\theta_{12}^2 + z_3^2\theta_{23}^2 +z_1 z_3 (\theta_{12}^2+\theta_{23}^2-\theta_{13}^2)\,,\\
\theta_{3\hat A}^2 &= z_1^2\theta_{13}^2 + z_2^2\theta_{23}^2 +z_1z_2 (\theta_{13}^2+\theta_{23}^2-\theta_{12}^2)\,.
\end{align}
The values of the three 1-subjettiness observables are then:
\begin{align}
\tau_1^{(0.5)} &= z_1\theta_{1\hat A}^{0.5}+z_2\theta_{2\hat A}^{0.5}+z_3\theta_{3\hat A}^{0.5}  \,,\\
\tau_1^{(1)} &= z_1\theta_{1\hat A}+z_2\theta_{2\hat A}+z_3\theta_{3\hat A}\,,\\
\tau_1^{(2)} &= z_1\theta_{1\hat A}^2+z_2\theta_{2\hat A}^2+z_3\theta_{3\hat A}^2 =z_1 z_2 \theta_{12}^2+z_1 z_3 \theta_{13}^2+z_2 z_3 \theta_{23}^2\,.
\end{align}
For $\tau_1^{(2)}$, the expression simplifies significantly in terms of the momentum fractions and pairwise angles. 

\section{Herwig Results}\label{app:herwig}

In this appendix, we present discrimination results for jets showered in Herwig 7.0.4 \cite{Bahr:2008pv,Bellm:2015jjp} from events generated in MadGraph.  The number of events showered in Herwig is about a factor of 10 fewer than that shown in the main body of the paper with Pythia, and so the neural network training is not as efficient.  Nevertheless, the conclusions drawn from this reduced Herwig sample are the same as from Pythia; namely, that observables sensitive to 4-body phase space saturate discrimination power.

On the sample of jets from $pp\to ZZ$, with one $Z$ decaying hadronically, and $pp\to Z+$ jet, we identify the same jets and measure the same collection of $N$-subjettiness observables as described in the main text.  These observables are then passed through the deep neural network as described in \Sec{sec:deeplearn}, with 390,000 events each for $pp\to ZZ$ and $pp\to Z+$ jet processes. These events were divided into 684,000 used for training, 76,000 for validation, and 20,000 for testing.

\begin{figure}
\begin{center}
\includegraphics[width=.6\textwidth]{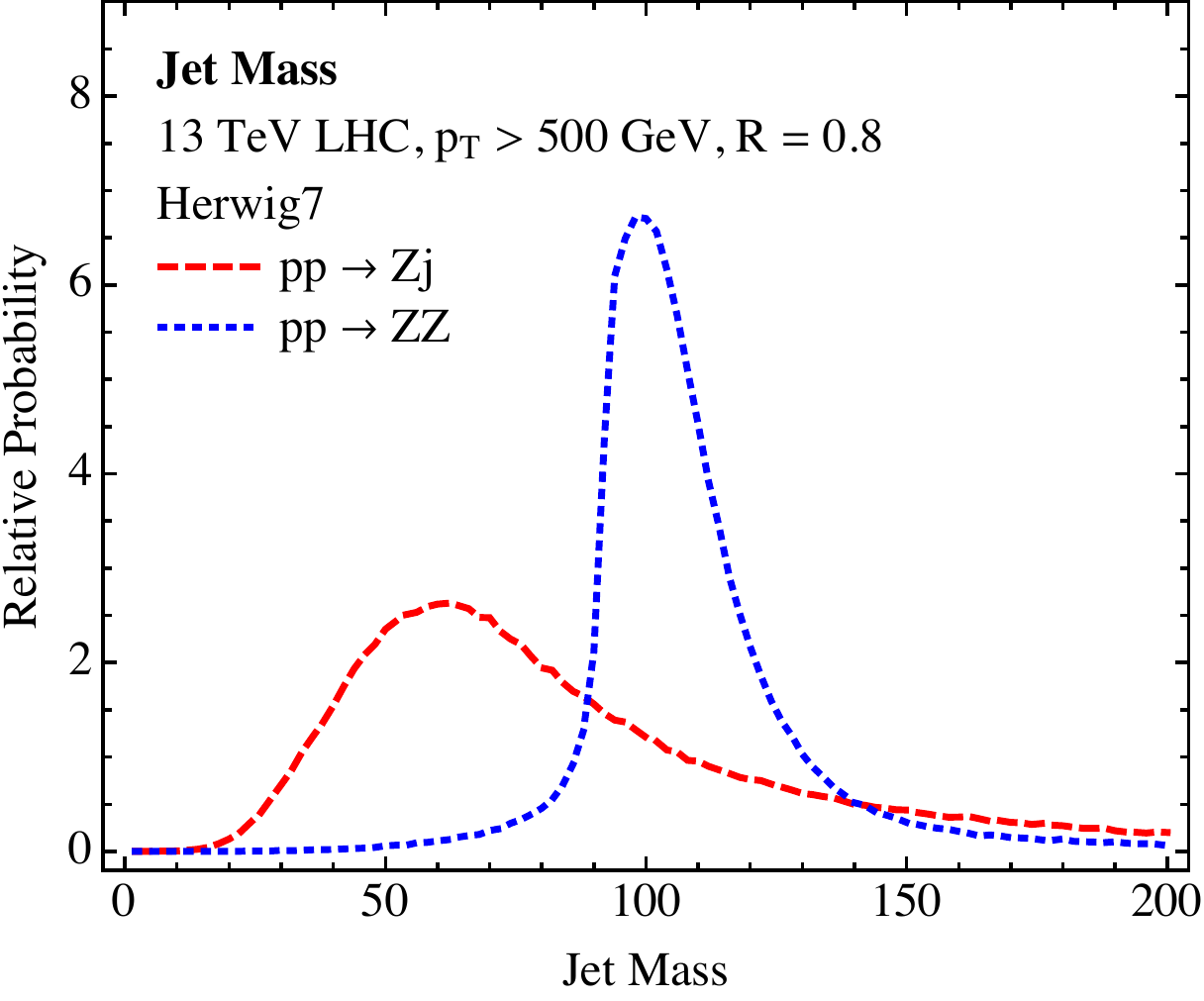}
\caption{
Distribution of the mass of the jet in $pp\to Zj$ (red dashed) and hadronically-decaying $Z$ boson in $pp\to ZZ$ (blue dotted) from the Herwig parton shower.  The minimum transverse momentum is 500 GeV, and the jets are found with the anti-$k_T$ algorithm with radius $R=0.8$.
}
\label{fig:massplot_h}
\end{center}
\end{figure}

\begin{figure}
\begin{center}
\subfloat[]{\label{fig:ht21}
\includegraphics[width=7cm]{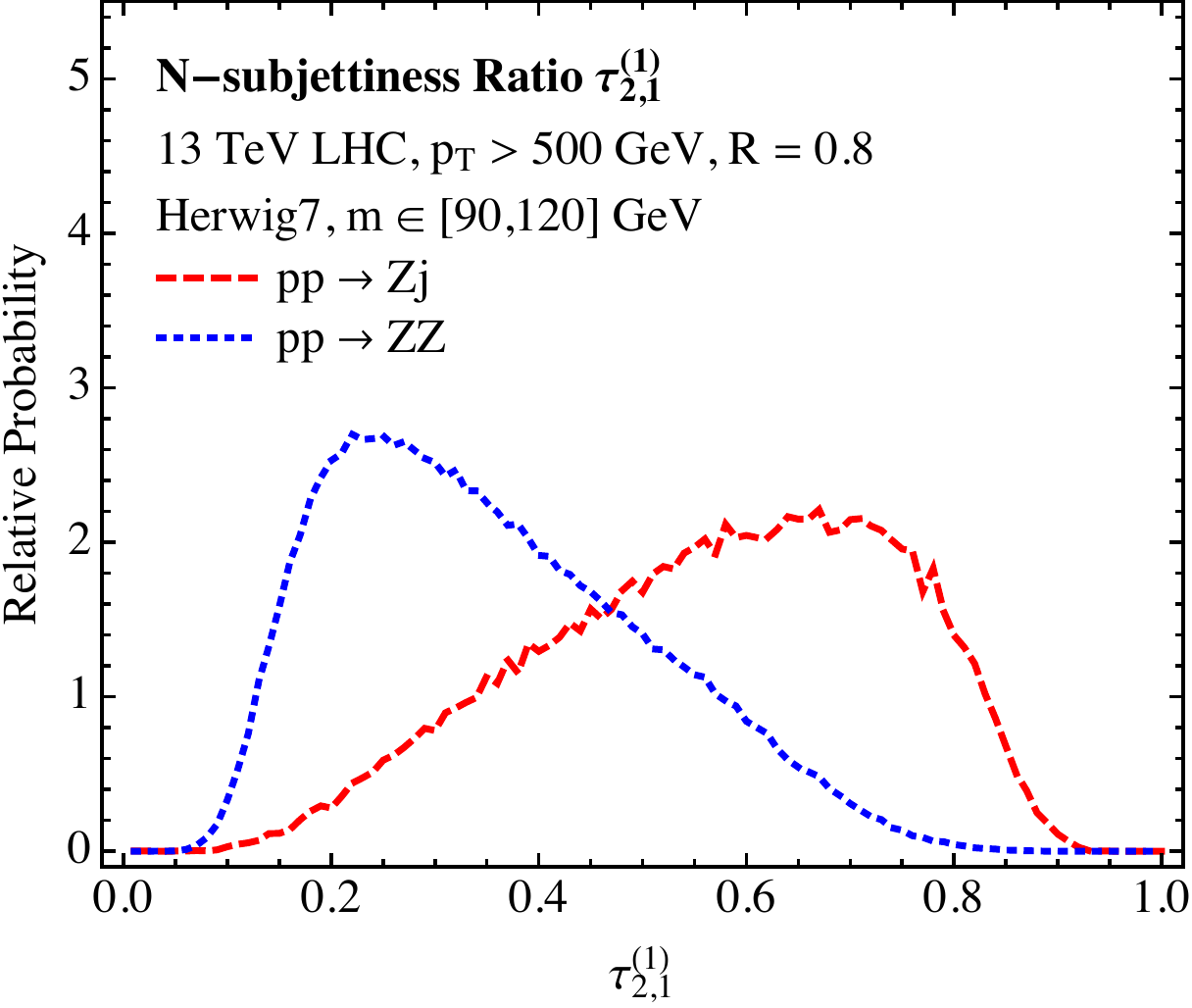}    
}\qquad
\subfloat[]{\label{fig:ht22}
\includegraphics[width=7cm]{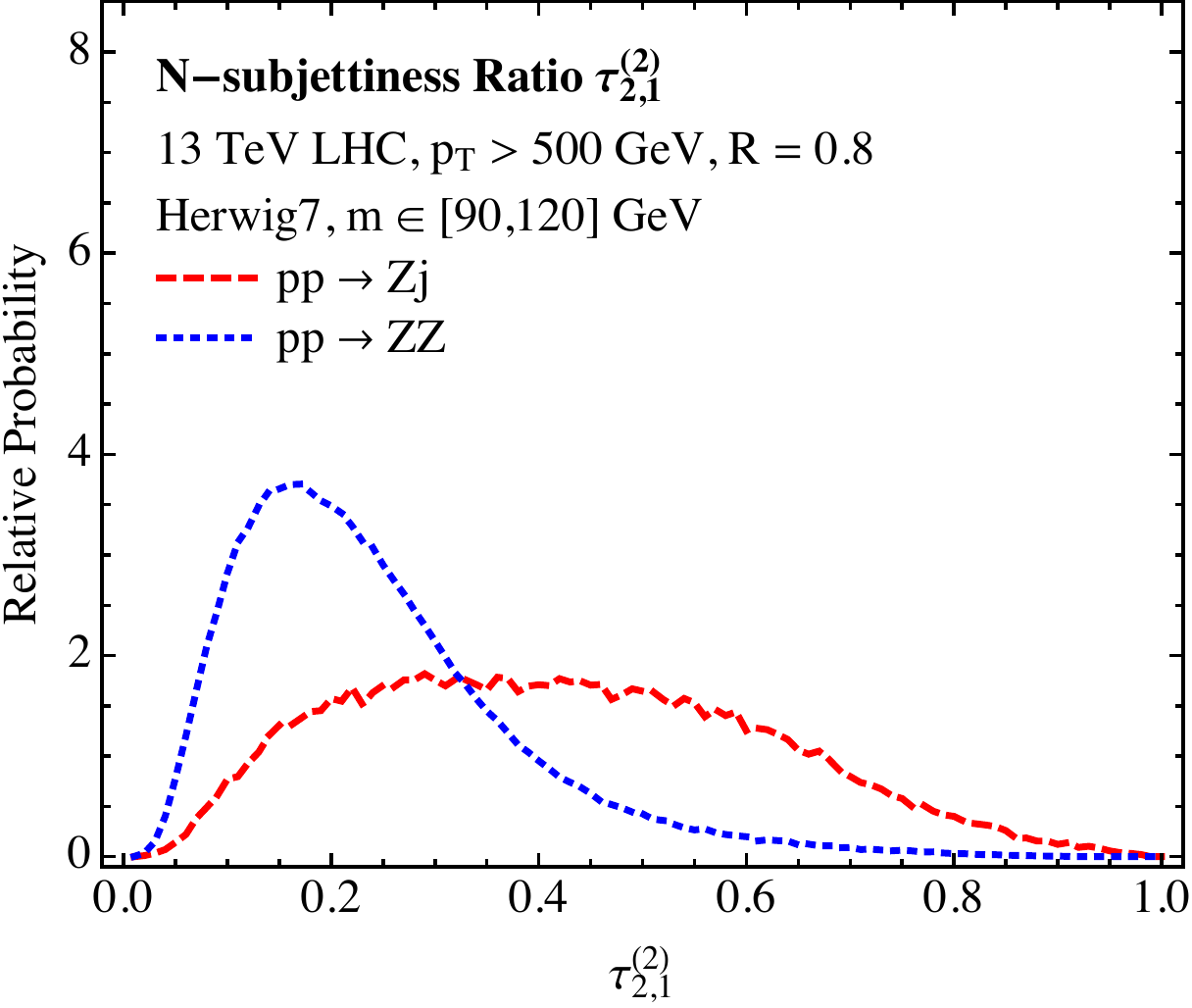}
}\\
\subfloat[]{\label{fig:hd21}
\includegraphics[width=7cm]{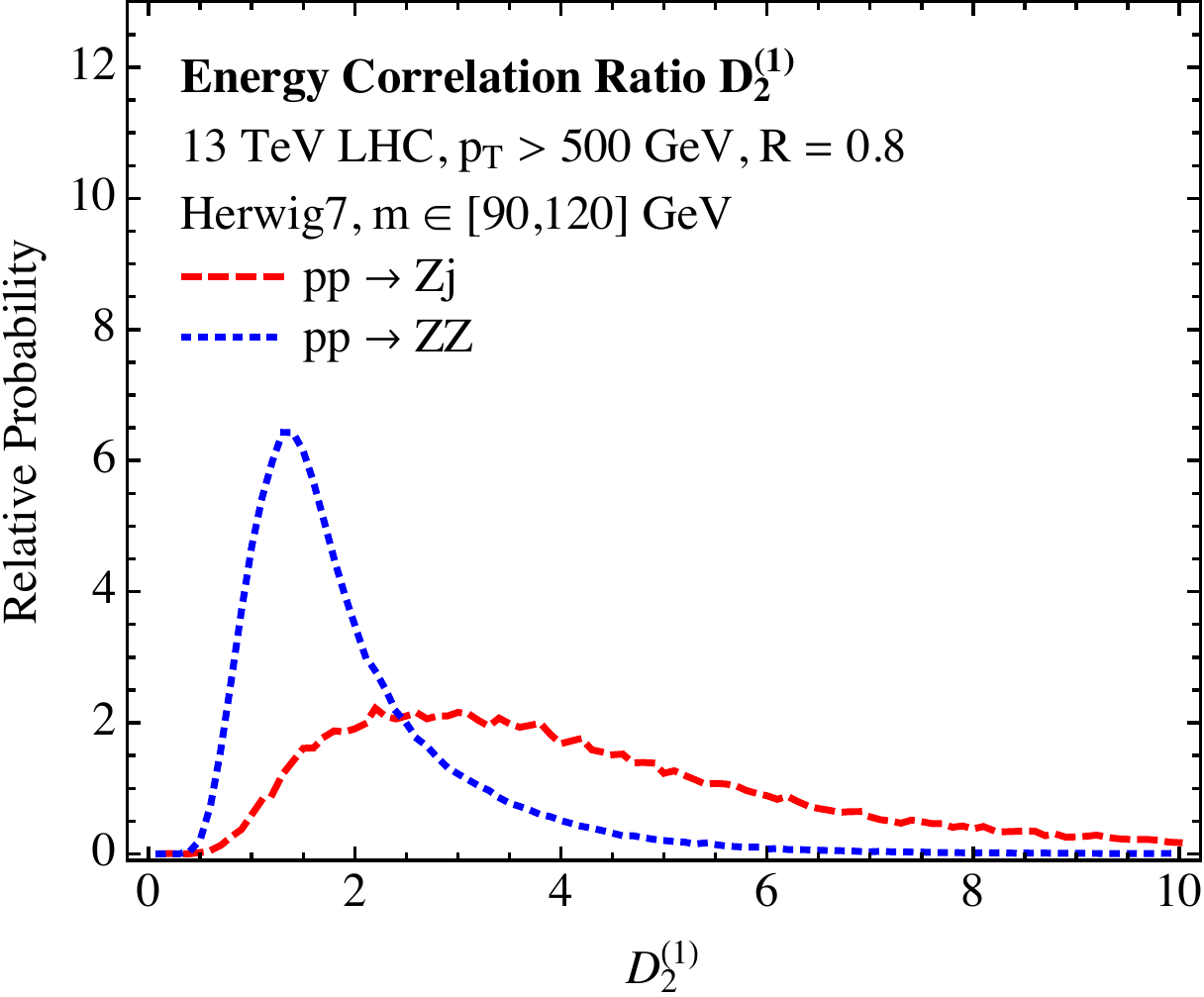}    
}\qquad
\subfloat[]{\label{fig:hd22}
\includegraphics[width=7cm]{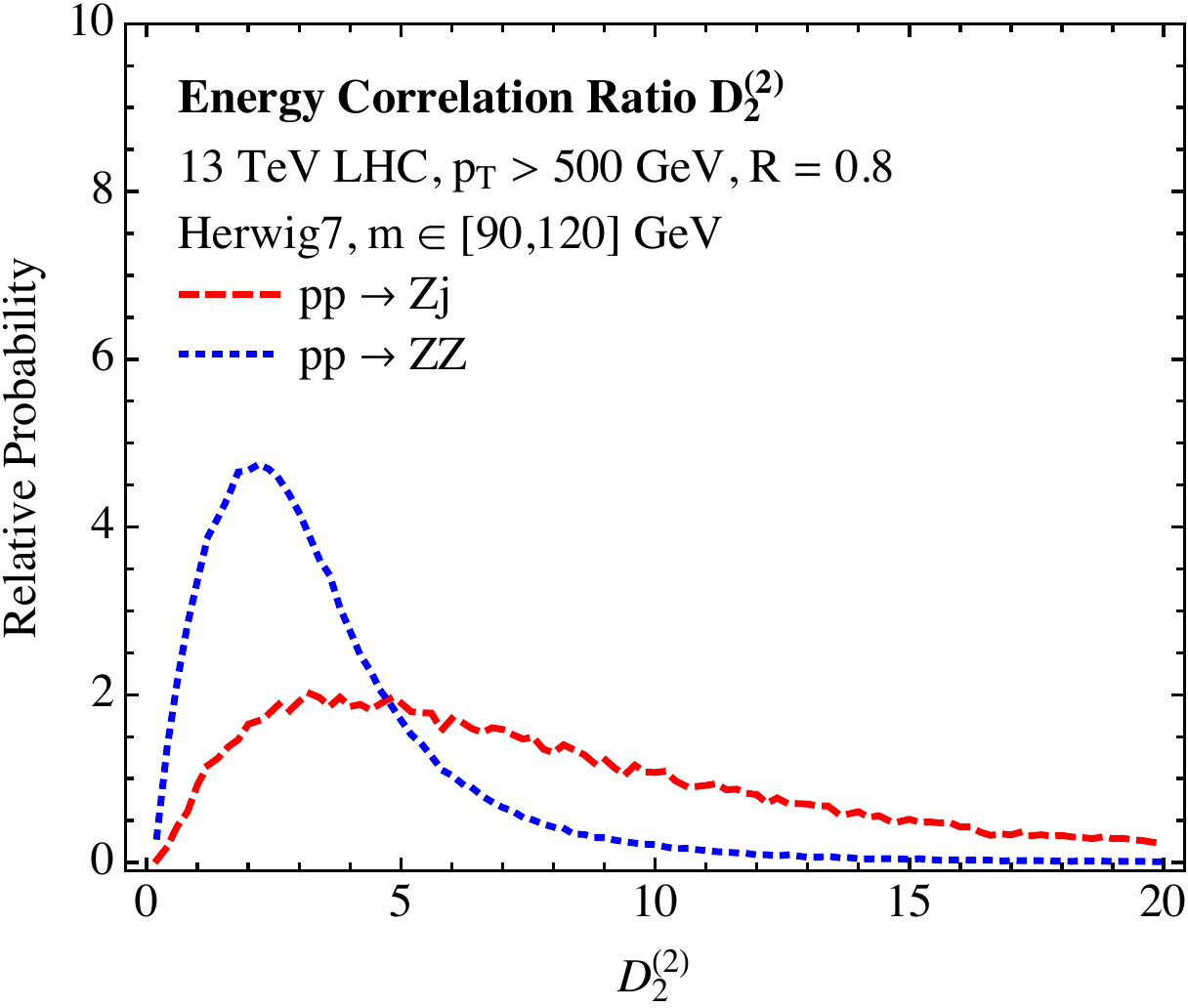}
}\\
\subfloat[]{\label{fig:hn21}
\includegraphics[width=7cm]{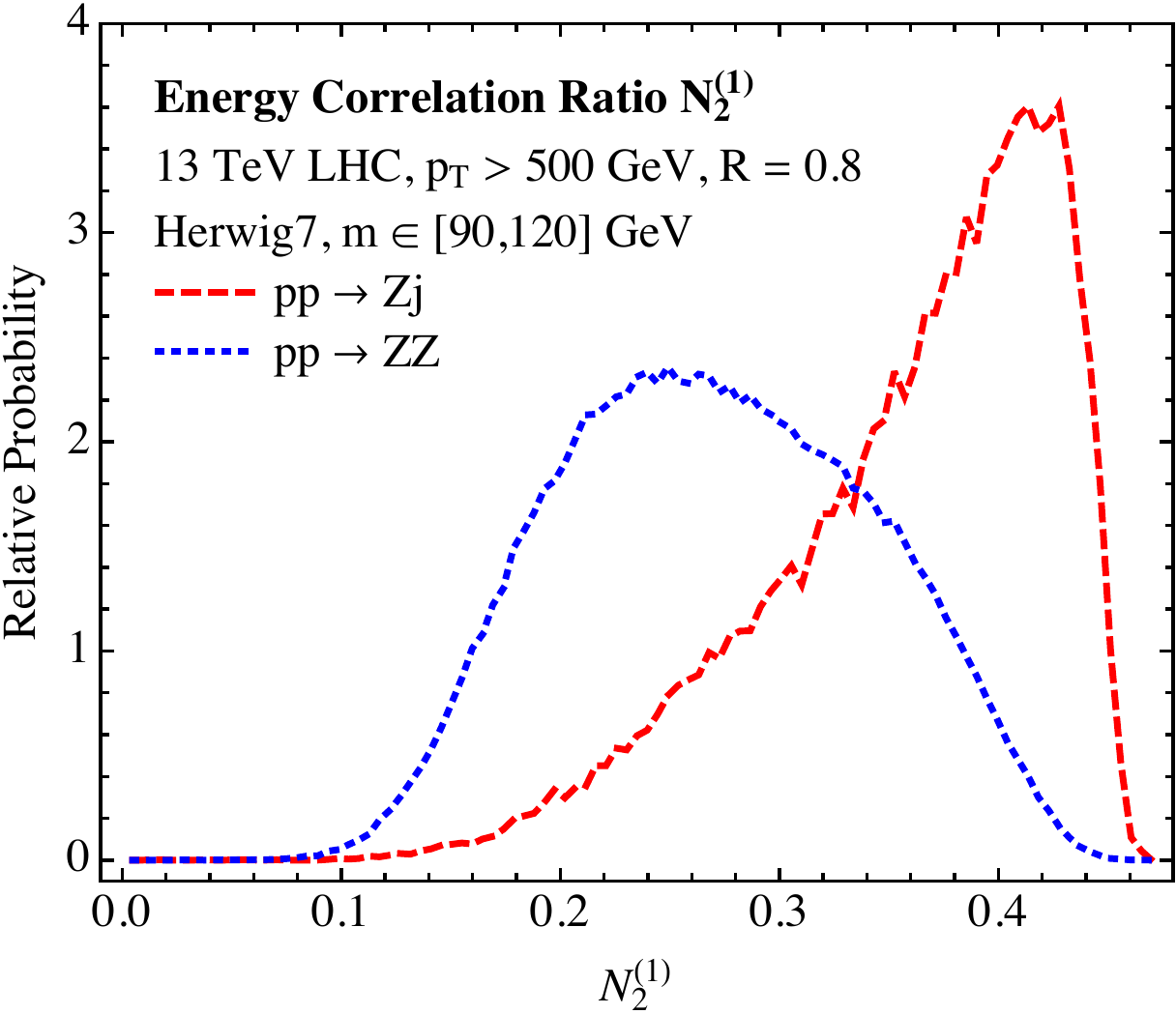}    
}\qquad
\subfloat[]{\label{fig:hn22}
\includegraphics[width=7cm]{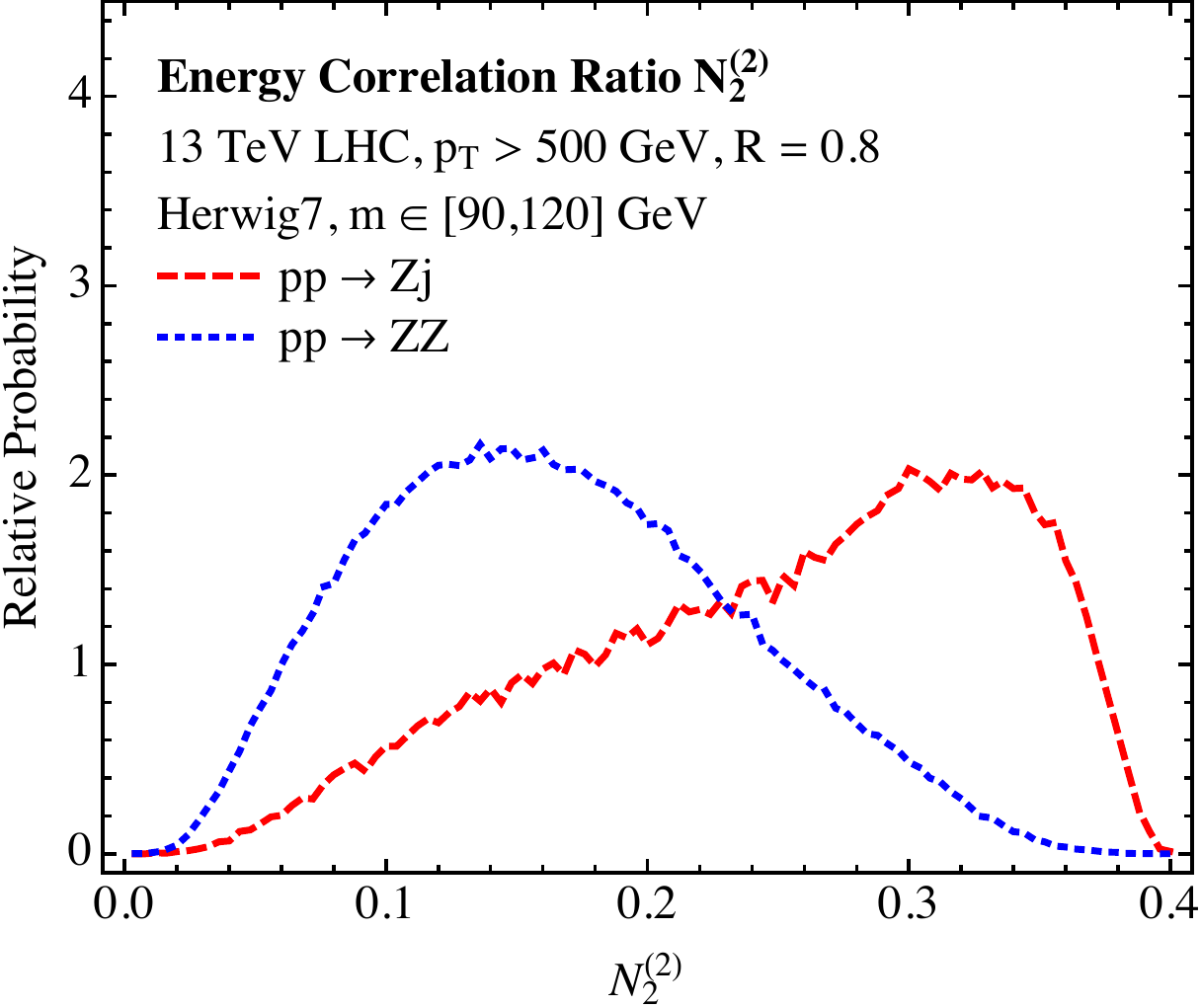}
}
\end{center}
\caption{
Distributions of various two-prong discrimination observables measured on the sample of jets showered with Herwig, on which a mass cut of $m\in[90,120]$ GeV has been placed.  From top to bottom are plotted signal (blue dotted) and background (red dashed) distributions of: $N$-subjettiness ratios $\tau_{2,1}^{(1)}$ (left) and $\tau_{2,1}^{(2)}$ (right), energy correlation function ratios $D_{2}^{(1)}$ (left) and $D_{2}^{(2)}$ (right), and $N_{2}^{(1)}$ (left) and $N_{2}^{(2)}$ (right).
}
\label{fig:obsdistros_h}
\end{figure}

In \Figs{fig:massplot_h}{fig:obsdistros_h}, we show validation plots on the jets showered with Herwig, to be compared with \Figs{fig:massplot}{fig:obsdistros} from Pythia.  The jet mass distribution in \Fig{fig:massplot_h} agrees qualitatively well with the corresponding plot from Pythia; though the Herwig sample seems to lack the small shoulder of the $Z$ boson mass distribution present in Pythia.  With a cut on the jet mass around the location of the $Z$ boson peak, we then measure the same selection of one- versus two-prong discriminant variables in the Herwig sample.  Again, good qualitative agreement is seem with Pythia, though the effects of finite statistics are much more evident.

\begin{figure}
\begin{center}
\includegraphics[width=.6\textwidth]{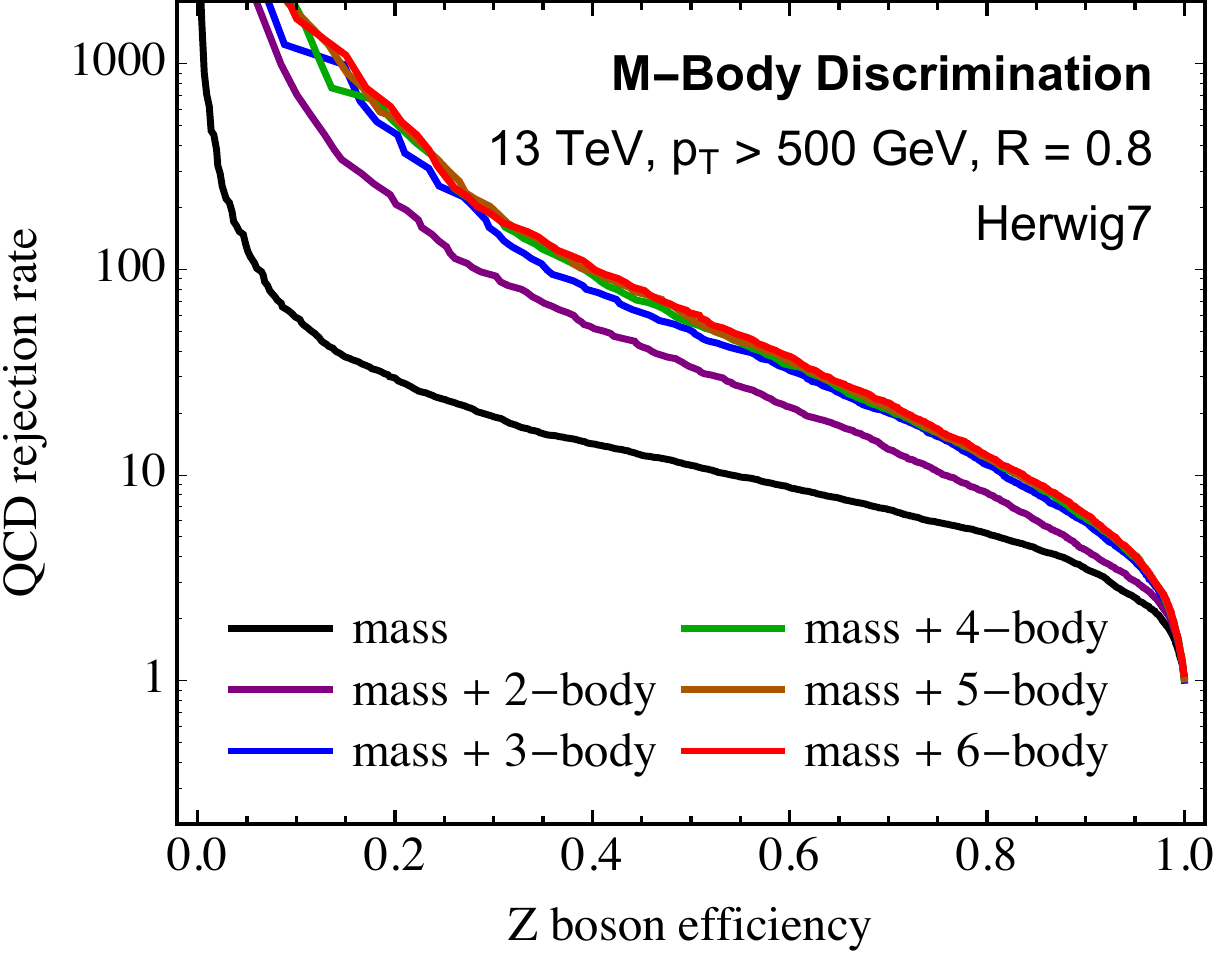}
\caption{
$Z$ boson jet efficiency vs.~QCD jet rejection rate plot generated by the deep neural network for jets showered in Herwig.  The different curves correspond to the mass plus collections of $N$-subjettiness observables that uniquely define $M$-body phase space.  Discrimination power is seen to saturate when 3- or 4-body phase space is resolved.
}
\label{fig:herwig}
\end{center}
\end{figure}

\Fig{fig:herwig} shows the signal efficiency vs.~background rejection rate for the collections of observables that resolve $M$-body phase space as determined by the neural network.  Just like in the Pythia samples, the discrimination power is observed to increase as more $N$-subjettiness observables are included.  The discrimination power is observed to saturate with observables that are sensitive to 3- or 4-body phase space.  This difference from when the Pythia events saturated could be due to the smaller jet sample size, though it could also be due to differences between the Pythia and Herwig parton showers.  It has been observed in numerous other studies \cite{Gallicchio:2012ez,Aad:2014gea,Larkoski:2014pca,Izaguirre:2014ira,Larkoski:2015kga,Larkoski:2014gra,Badger:2016bpw,Aad:2016oit,Gras:2017jty} that the discrimination performance differs significantly between jets showered in Pythia versus Herwig.  The exact reason for the discrepancy is beyond this paper, but the existence of a saturation point also in Herwig demonstrates that there is only a very limited amount of information in the jet for discrimination.

\section{Results with Other Architectures}\label{app:morearchs}

In this appendix, we show discrimination results for a neural network with one more hidden layer than the network studied in the body of the paper, as well as the output of a boosted decision tree.

\subsection{A Deeper Neural Network}

The neural network used in this appendix is identical to the network studied in the body of the paper, except with the addition of another layer.  Immediately after the input layer, we have included an additional Dense layer of 1000 nodes, with a Dropout regularization of 0.2.  The typical number of training epochs of this new neural network was about 50 for each collection of observables.

\begin{figure}
\begin{center}
\includegraphics[width=.6\textwidth]{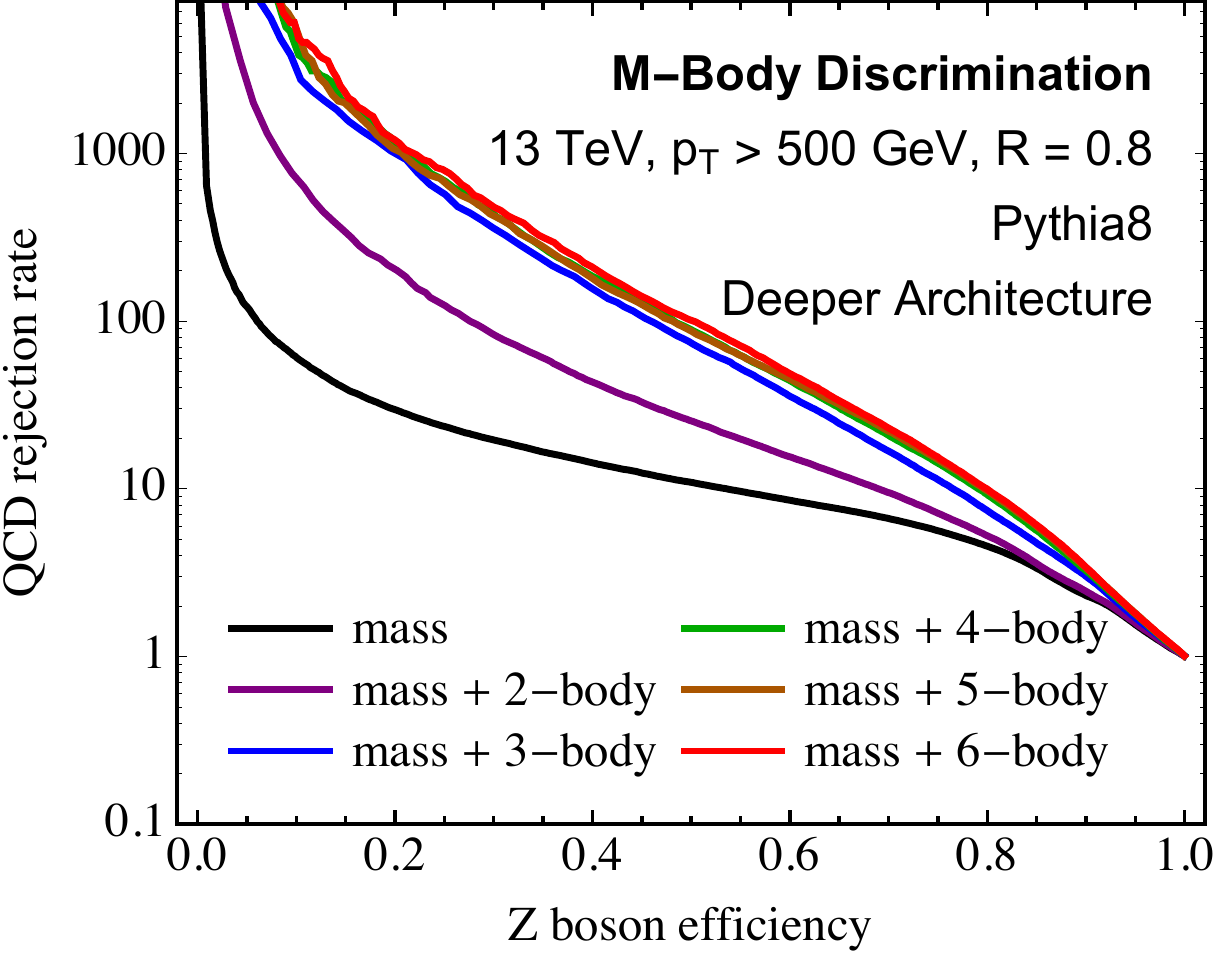}
\caption{
$Z$ boson jet efficiency vs.~QCD jet rejection rate plot as generated by the deeper neural network for events showered in Pythia.  The different curves correspond to the mass plus collections of observables that uniquely define $M$-body phase space.  Discrimination power is seen to saturate when 4-body phase space is resolved.
}
\label{fig:deeperplot}
\end{center}
\end{figure}

\begin{figure}
\begin{center}
\includegraphics[width=.6\textwidth]{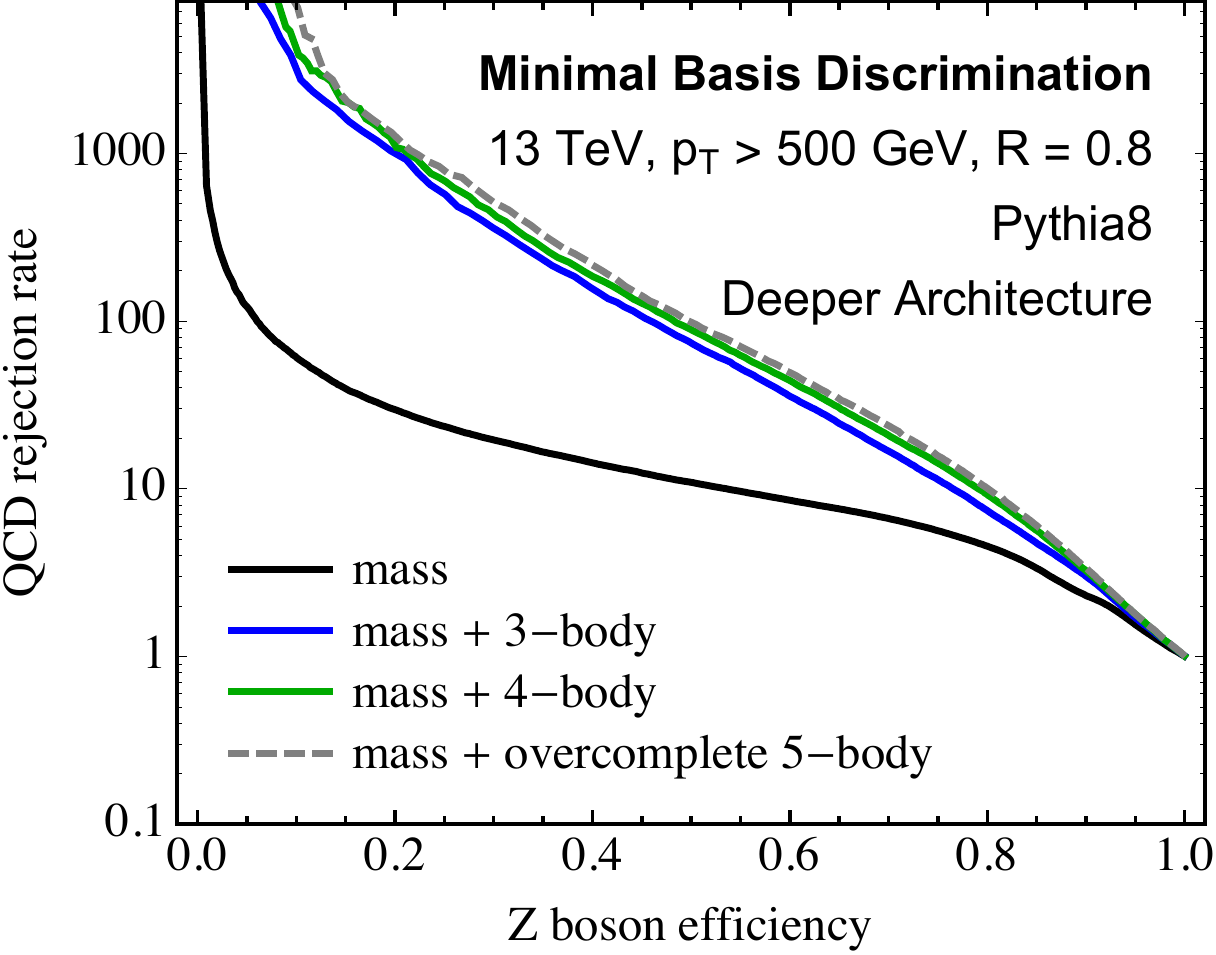}
\caption{
Signal efficiency versus background rejection rate for jet mass plus the overcomplete basis of observables that are sensitive to 5-body phase space described in the text, as determined by the deeper neural network.  For comparison, we also include the signal efficiency versus background rejection rate for jet mass, jet mass plus minimal 3-body phase space observables, and jet mass plus the minimal 4-body phase space observables.
}
\label{fig:deeperoverplot}
\end{center}
\end{figure}

We show the discrimination performance as identified by this network in \Figs{fig:deeperplot}{fig:deeperoverplot}.  In \Fig{fig:deeperplot}, we show the discrimination power as more observables are added to resolve higher-body phase space.  As with the other studies in this paper, we see that the discrimination power is saturated when 4-body phase space is resolved.  Additionally in \Fig{fig:deeperoverplot}, we compare the discrimination power of 3- and 4-body phase space observables to the overcomplete 5-body phase space observables described in \Sec{sec:deeplearn}.  The overcomplete basis of observables is observed to be only very slightly better than 4-body phase space basis, suggesting that essentially all useful discrimination information has been extracted.

\subsection{Boosted Decision Tree}

\begin{figure}
\begin{center}
\includegraphics[width=.6\textwidth]{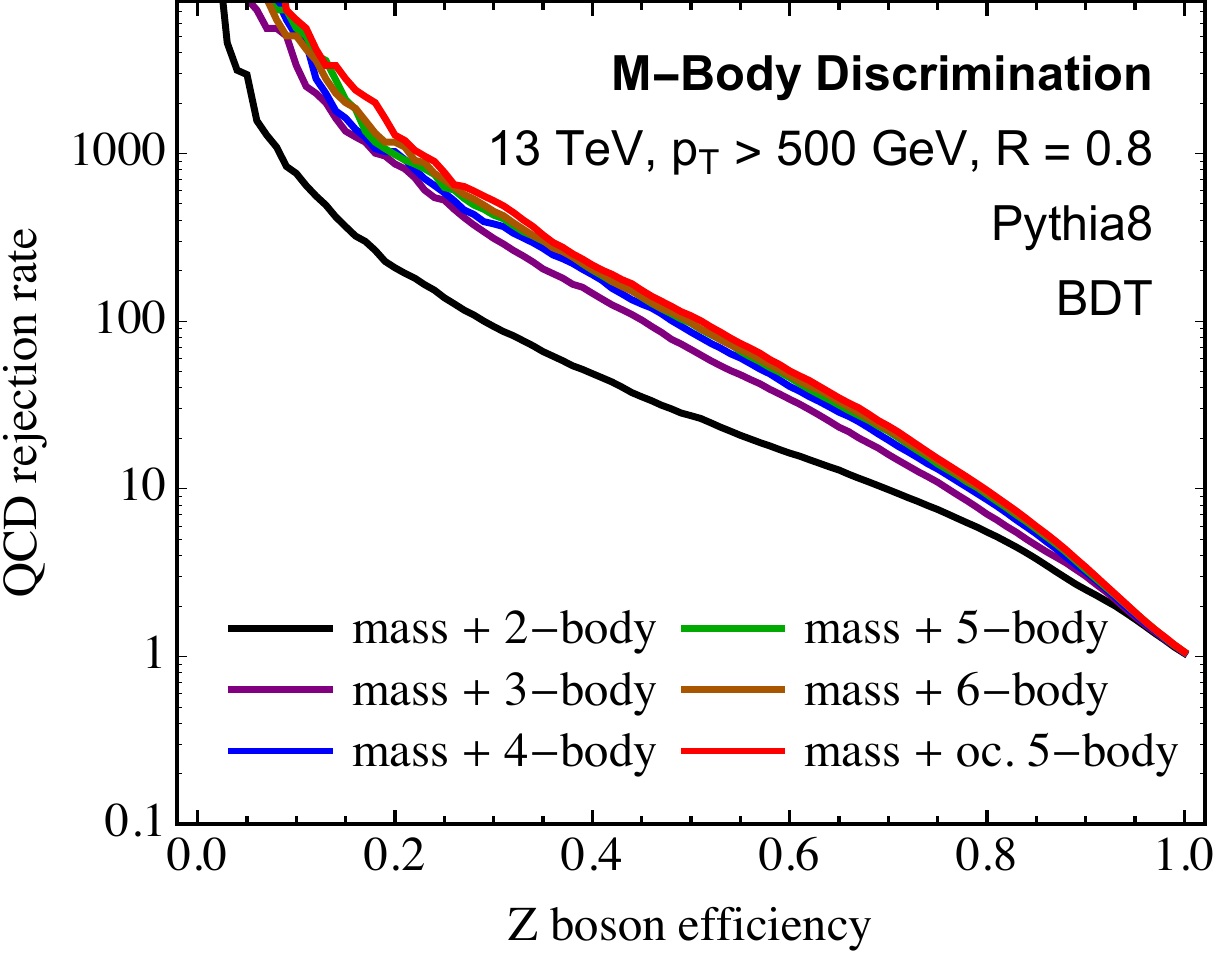}
\caption{
$Z$ boson jet efficiency vs.~QCD jet rejection rate plot as generated by the boosted decision tree for events showered in Pythia.  The different curves correspond to the mass plus collections of observables that uniquely define $M$-body phase space.  Discrimination power is seen to saturate when 4-body phase space is resolved.  In this plot, we also include the overcomplete 5-body phase space collection of observables, labeled ``oc.~5-body''.
}
\label{fig:bdtplot}
\end{center}
\end{figure}

Because our observable basis is quite small, we can input them to a boosted decision tree to evaluate the discrimination power.  We used ROOT's TMVA package \cite{Hocker:2007ht,Speckmayer:2010zz} to train and test the boosted decision trees.  Each collection of phase space observables studied elsewhere in this paper were input to the boosted decision trees, and forests of 2500 trees were used.  We also trained on forests of 850 trees, and observed no significant improvement in discrimination power in extending to forests of 2500 trees, suggesting that the boosted decision trees are extracting all the information that they can.  The results of the boosted decision trees are shown in \Fig{fig:bdtplot}.  These results are again consistent with what we found earlier; namely, that discrimination power is observed to saturate once 4-body phase space is resolved.

\bibliography{machinelearning}

\providecommand{\href}[2]{#2}\begingroup\raggedright\begin{thebibliography}{10}

\bibitem{Adams:2015hiv}
D.~Adams et~al., {\it {Towards an Understanding of the Correlations in Jet
  Substructure}},  {\em Eur. Phys. J.} {\bf C75} (2015), no.~9 409,
  [\href{http://arxiv.org/abs/1504.00679}{{\tt arXiv:1504.00679}}].

\bibitem{Altheimer:2013yza}
A.~Altheimer et~al., {\it {Boosted objects and jet substructure at the LHC.
  Report of BOOST2012, held at IFIC Valencia, 23rd-27th of July 2012}},  {\em
  Eur. Phys. J.} {\bf C74} (2014), no.~3 2792,
  [\href{http://arxiv.org/abs/1311.2708}{{\tt arXiv:1311.2708}}].

\bibitem{Altheimer:2012mn}
A.~Altheimer et~al., {\it {Jet Substructure at the Tevatron and LHC: New
  results, new tools, new benchmarks}},  {\em J. Phys.} {\bf G39} (2012)
  063001, [\href{http://arxiv.org/abs/1201.0008}{{\tt arXiv:1201.0008}}].

\bibitem{Abdesselam:2010pt}
A.~Abdesselam et~al., {\it {Boosted objects: A Probe of beyond the Standard
  Model physics}},  {\em Eur. Phys. J.} {\bf C71} (2011) 1661,
  [\href{http://arxiv.org/abs/1012.5412}{{\tt arXiv:1012.5412}}].

\bibitem{Tkachov:1995kk}
F.~V. Tkachov, {\it {Measuring multi - jet structure of hadronic energy flow or
  What is a jet?}},  {\em Int. J. Mod. Phys.} {\bf A12} (1997) 5411--5529,
  [\href{http://arxiv.org/abs/hep-ph/9601308}{{\tt hep-ph/9601308}}].

\bibitem{Sveshnikov:1995vi}
N.~A. Sveshnikov and F.~V. Tkachov, {\it {Jets and quantum field theory}},
  {\em Phys. Lett.} {\bf B382} (1996) 403--408,
  [\href{http://arxiv.org/abs/hep-ph/9512370}{{\tt hep-ph/9512370}}].

\bibitem{Cherzor:1997ak}
P.~S. Cherzor and N.~A. Sveshnikov, {\it {Jet observables and energy momentum
  tensor}},  in {\em {Quantum field theory and high-energy physics.
  Proceedings, Workshop, QFTHEP'97, Samara, Russia, September 4-10, 1997}},
  pp.~402--407, 1997.
\newblock \href{http://arxiv.org/abs/hep-ph/9710349}{{\tt hep-ph/9710349}}.

\bibitem{Tkachov:1999py}
F.~V. Tkachov, {\it {A Theory of jet definition}},  {\em Int. J. Mod. Phys.}
  {\bf A17} (2002) 2783--2884, [\href{http://arxiv.org/abs/hep-ph/9901444}{{\tt
  hep-ph/9901444}}].

\bibitem{Cogan:2014oua}
J.~Cogan, M.~Kagan, E.~Strauss, and A.~Schwarztman, {\it {Jet-Images: Computer
  Vision Inspired Techniques for Jet Tagging}},  {\em JHEP} {\bf 02} (2015)
  118, [\href{http://arxiv.org/abs/1407.5675}{{\tt arXiv:1407.5675}}].

\bibitem{Almeida:2015jua}
L.~G. Almeida, M.~Backovic, M.~Cliche, S.~J. Lee, and M.~Perelstein, {\it
  {Playing Tag with ANN: Boosted Top Identification with Pattern Recognition}},
   {\em JHEP} {\bf 07} (2015) 086, [\href{http://arxiv.org/abs/1501.05968}{{\tt
  arXiv:1501.05968}}].

\bibitem{deOliveira:2015xxd}
L.~de~Oliveira, M.~Kagan, L.~Mackey, B.~Nachman, and A.~Schwartzman, {\it
  {Jet-images ? deep learning edition}},  {\em JHEP} {\bf 07} (2016) 069,
  [\href{http://arxiv.org/abs/1511.05190}{{\tt arXiv:1511.05190}}].

\bibitem{Baldi:2016fql}
P.~Baldi, K.~Bauer, C.~Eng, P.~Sadowski, and D.~Whiteson, {\it {Jet
  Substructure Classification in High-Energy Physics with Deep Neural
  Networks}},  {\em Phys. Rev.} {\bf D93} (2016), no.~9 094034,
  [\href{http://arxiv.org/abs/1603.09349}{{\tt arXiv:1603.09349}}].

\bibitem{Guest:2016iqz}
D.~Guest, J.~Collado, P.~Baldi, S.-C. Hsu, G.~Urban, and D.~Whiteson, {\it {Jet
  Flavor Classification in High-Energy Physics with Deep Neural Networks}},
  {\em Phys. Rev.} {\bf D94} (2016), no.~11 112002,
  [\href{http://arxiv.org/abs/1607.08633}{{\tt arXiv:1607.08633}}].

\bibitem{Conway:2016caq}
J.~S. Conway, R.~Bhaskar, R.~D. Erbacher, and J.~Pilot, {\it {Identification of
  High-Momentum Top Quarks, Higgs Bosons, and W and Z Bosons Using Boosted
  Event Shapes}},  {\em Phys. Rev.} {\bf D94} (2016), no.~9 094027,
  [\href{http://arxiv.org/abs/1606.06859}{{\tt arXiv:1606.06859}}].

\bibitem{Barnard:2016qma}
J.~Barnard, E.~N. Dawe, M.~J. Dolan, and N.~Rajcic, {\it {Parton Shower
  Uncertainties in Jet Substructure Analyses with Deep Neural Networks}},  {\em
  Phys. Rev.} {\bf D95} (2017), no.~1 014018,
  [\href{http://arxiv.org/abs/1609.00607}{{\tt arXiv:1609.00607}}].

\bibitem{Komiske:2016rsd}
P.~T. Komiske, E.~M. Metodiev, and M.~D. Schwartz, {\it {Deep learning in
  color: towards automated quark/gluon jet discrimination}},  {\em JHEP} {\bf
  01} (2017) 110, [\href{http://arxiv.org/abs/1612.01551}{{\tt
  arXiv:1612.01551}}].

\bibitem{deOliveira:2017pjk}
L.~de~Oliveira, M.~Paganini, and B.~Nachman, {\it {Learning Particle Physics by
  Example: Location-Aware Generative Adversarial Networks for Physics
  Synthesis}},  \href{http://arxiv.org/abs/1701.05927}{{\tt arXiv:1701.05927}}.

\bibitem{Kasieczka:2017nvn}
G.~Kasieczka, T.~Plehn, M.~Russell, and T.~Schell, {\it {Deep-learning Top
  Taggers or The End of QCD?}},  \href{http://arxiv.org/abs/1701.08784}{{\tt
  arXiv:1701.08784}}.

\bibitem{Louppe:2017ipp}
G.~Louppe, K.~Cho, C.~Becot, and K.~Cranmer, {\it {QCD-Aware Recursive Neural
  Networks for Jet Physics}},  \href{http://arxiv.org/abs/1702.00748}{{\tt
  arXiv:1702.00748}}.

\bibitem{Dery:2017fap}
L.~M. Dery, B.~Nachman, F.~Rubbo, and A.~Schwartzman, {\it {Weakly Supervised
  Classification in High Energy Physics}},
  \href{http://arxiv.org/abs/1702.00414}{{\tt arXiv:1702.00414}}.

\bibitem{Pearkes:2017hku}
J.~Pearkes, W.~Fedorko, A.~Lister, and C.~Gay, {\it {Jet Constituents for Deep
  Neural Network Based Top Quark Tagging}},
  \href{http://arxiv.org/abs/1704.02124}{{\tt arXiv:1704.02124}}.

\bibitem{Larkoski:2013paa}
A.~J. Larkoski and J.~Thaler, {\it {Unsafe but Calculable: Ratios of
  Angularities in Perturbative QCD}},  {\em JHEP} {\bf 09} (2013) 137,
  [\href{http://arxiv.org/abs/1307.1699}{{\tt arXiv:1307.1699}}].

\bibitem{Larkoski:2015lea}
A.~J. Larkoski, S.~Marzani, and J.~Thaler, {\it {Sudakov Safety in Perturbative
  QCD}},  {\em Phys. Rev.} {\bf D91} (2015), no.~11 111501,
  [\href{http://arxiv.org/abs/1502.01719}{{\tt arXiv:1502.01719}}].

\bibitem{Stewart:2010tn}
I.~W. Stewart, F.~J. Tackmann, and W.~J. Waalewijn, {\it {N-Jettiness: An
  Inclusive Event Shape to Veto Jets}},  {\em Phys. Rev. Lett.} {\bf 105}
  (2010) 092002, [\href{http://arxiv.org/abs/1004.2489}{{\tt
  arXiv:1004.2489}}].

\bibitem{Thaler:2010tr}
J.~Thaler and K.~Van~Tilburg, {\it {Identifying Boosted Objects with
  N-subjettiness}},  {\em JHEP} {\bf 03} (2011) 015,
  [\href{http://arxiv.org/abs/1011.2268}{{\tt arXiv:1011.2268}}].

\bibitem{Thaler:2011gf}
J.~Thaler and K.~Van~Tilburg, {\it {Maximizing Boosted Top Identification by
  Minimizing N-subjettiness}},  {\em JHEP} {\bf 02} (2012) 093,
  [\href{http://arxiv.org/abs/1108.2701}{{\tt arXiv:1108.2701}}].

\bibitem{Larkoski:2013eya}
A.~J. Larkoski, G.~P. Salam, and J.~Thaler, {\it {Energy Correlation Functions
  for Jet Substructure}},  {\em JHEP} {\bf 06} (2013) 108,
  [\href{http://arxiv.org/abs/1305.0007}{{\tt arXiv:1305.0007}}].

\bibitem{Moult:2016cvt}
I.~Moult, L.~Necib, and J.~Thaler, {\it {New Angles on Energy Correlation
  Functions}},  {\em JHEP} {\bf 12} (2016) 153,
  [\href{http://arxiv.org/abs/1609.07483}{{\tt arXiv:1609.07483}}].

\bibitem{Catani:1993hr}
S.~Catani, Y.~L. Dokshitzer, M.~H. Seymour, and B.~R. Webber, {\it
  {Longitudinally invariant $K_t$ clustering algorithms for hadron hadron
  collisions}},  {\em Nucl. Phys.} {\bf B406} (1993) 187--224.

\bibitem{Ellis:1993tq}
S.~D. Ellis and D.~E. Soper, {\it {Successive combination jet algorithm for
  hadron collisions}},  {\em Phys. Rev.} {\bf D48} (1993) 3160--3166,
  [\href{http://arxiv.org/abs/hep-ph/9305266}{{\tt hep-ph/9305266}}].

\bibitem{Blazey:2000qt}
G.~C. Blazey et~al., {\it {Run II jet physics}},  in {\em {QCD and weak boson
  physics in Run II. Proceedings, Batavia, USA, March 4-6, June 3-4, November
  4-6, 1999}}, pp.~47--77, 2000.
\newblock \href{http://arxiv.org/abs/hep-ex/0005012}{{\tt hep-ex/0005012}}.

\bibitem{Bertolini:2013iqa}
D.~Bertolini, T.~Chan, and J.~Thaler, {\it {Jet Observables Without Jet
  Algorithms}},  {\em JHEP} {\bf 04} (2014) 013,
  [\href{http://arxiv.org/abs/1310.7584}{{\tt arXiv:1310.7584}}].

\bibitem{Larkoski:2014uqa}
A.~J. Larkoski, D.~Neill, and J.~Thaler, {\it {Jet Shapes with the Broadening
  Axis}},  {\em JHEP} {\bf 04} (2014) 017,
  [\href{http://arxiv.org/abs/1401.2158}{{\tt arXiv:1401.2158}}].

\bibitem{Larkoski:2014bia}
A.~J. Larkoski and J.~Thaler, {\it {Aspects of jets at 100 TeV}},  {\em Phys.
  Rev.} {\bf D90} (2014), no.~3 034010,
  [\href{http://arxiv.org/abs/1406.7011}{{\tt arXiv:1406.7011}}].

\bibitem{Alwall:2014hca}
J.~Alwall, R.~Frederix, S.~Frixione, V.~Hirschi, F.~Maltoni, O.~Mattelaer,
  H.~S. Shao, T.~Stelzer, P.~Torrielli, and M.~Zaro, {\it {The automated
  computation of tree-level and next-to-leading order differential cross
  sections, and their matching to parton shower simulations}},  {\em JHEP} {\bf
  07} (2014) 079, [\href{http://arxiv.org/abs/1405.0301}{{\tt
  arXiv:1405.0301}}].

\bibitem{Sjostrand:2006za}
T.~Sjostrand, S.~Mrenna, and P.~Z. Skands, {\it {PYTHIA 6.4 Physics and
  Manual}},  {\em JHEP} {\bf 05} (2006) 026,
  [\href{http://arxiv.org/abs/hep-ph/0603175}{{\tt hep-ph/0603175}}].

\bibitem{Sjostrand:2014zea}
T.~Sj�strand, S.~Ask, J.~R. Christiansen, R.~Corke, N.~Desai, P.~Ilten,
  S.~Mrenna, S.~Prestel, C.~O. Rasmussen, and P.~Z. Skands, {\it {An
  Introduction to PYTHIA 8.2}},  {\em Comput. Phys. Commun.} {\bf 191} (2015)
  159--177, [\href{http://arxiv.org/abs/1410.3012}{{\tt arXiv:1410.3012}}].

\bibitem{Bahr:2008pv}
M.~Bahr et~al., {\it {Herwig++ Physics and Manual}},  {\em Eur. Phys. J.} {\bf
  C58} (2008) 639--707, [\href{http://arxiv.org/abs/0803.0883}{{\tt
  arXiv:0803.0883}}].

\bibitem{Bellm:2015jjp}
J.~Bellm et~al., {\it {Herwig 7.0/Herwig++ 3.0 release note}},  {\em Eur. Phys.
  J.} {\bf C76} (2016), no.~4 196, [\href{http://arxiv.org/abs/1512.01178}{{\tt
  arXiv:1512.01178}}].

\bibitem{Cacciari:2011ma}
M.~Cacciari, G.~P. Salam, and G.~Soyez, {\it {FastJet User Manual}},  {\em Eur.
  Phys. J.} {\bf C72} (2012) 1896, [\href{http://arxiv.org/abs/1111.6097}{{\tt
  arXiv:1111.6097}}].

\bibitem{Cacciari:2005hq}
M.~Cacciari and G.~P. Salam, {\it {Dispelling the $N^{3}$ myth for the $k_t$
  jet-finder}},  {\em Phys. Lett.} {\bf B641} (2006) 57--61,
  [\href{http://arxiv.org/abs/hep-ph/0512210}{{\tt hep-ph/0512210}}].

\bibitem{Cacciari:2008gp}
M.~Cacciari, G.~P. Salam, and G.~Soyez, {\it {The Anti-k(t) jet clustering
  algorithm}},  {\em JHEP} {\bf 04} (2008) 063,
  [\href{http://arxiv.org/abs/0802.1189}{{\tt arXiv:0802.1189}}].

\bibitem{Larkoski:2014gra}
A.~J. Larkoski, I.~Moult, and D.~Neill, {\it {Power Counting to Better Jet
  Observables}},  {\em JHEP} {\bf 12} (2014) 009,
  [\href{http://arxiv.org/abs/1409.6298}{{\tt arXiv:1409.6298}}].

\bibitem{hdf5}
{The HDF Group}, {\it {Hierarchical Data Format, version 5}},  1997-NNNN.
\newblock http://www.hdfgroup.org/HDF5/.

\bibitem{chollet2015keras}
F.~Chollet, ``Keras.'' \url{https://github.com/fchollet/keras}, 2015.

\bibitem{scikit-learn}
{Pedregosa, F., Varoquaux, G., Gramfort, A., Michel, V., Thirion, B., Grisel,
  O., Blondel, M., Prettenhofer, P., Weiss, R., Dubourg, V., Vanderplas, J.,
  Passos, A., Cournapeau, D., Brucher, M., Perrot and M., Duchesnay, E.}, {\it
  {Scikit-learn: Machine Learning in Python}},  {\em {J. Mach. Learn. Res.}}
  {\bf 12} (2011) 2825--2830.

\bibitem{dropout}
N.~Srivastava, G.~Hinton, A.~Krizhevsky, I.~Sutskever, and R.~Salakhutdinov,
  {\it Dropout: A simple way to prevent neural networks from overfitting},
  {\em J. Mach. Learn. Res.} {\bf 15} (Jan., 2014) 1929--1958.

\bibitem{conf/icml/NairH10}
V.~Nair and G.~E. Hinton, {\it Rectified linear units improve restricted
  boltzmann machines.},  in {\em ICML} (J.~Fürnkranz and T.~Joachims, eds.),
  pp.~807--814, Omnipress, 2010.

\bibitem{DBLP:journals/corr/KingmaB14}
D.~P. Kingma and J.~Ba, {\it Adam: {A} method for stochastic optimization},
  {\em CoRR} {\bf abs/1412.6980} (2014).

\bibitem{Larkoski:2015zka}
A.~J. Larkoski, I.~Moult, and D.~Neill, {\it {Non-Global Logarithms,
  Factorization, and the Soft Substructure of Jets}},  {\em JHEP} {\bf 09}
  (2015) 143, [\href{http://arxiv.org/abs/1501.04596}{{\tt arXiv:1501.04596}}].

\bibitem{Larkoski:2014tva}
A.~J. Larkoski, I.~Moult, and D.~Neill, {\it {Toward Multi-Differential Cross
  Sections: Measuring Two Angularities on a Single Jet}},  {\em JHEP} {\bf 09}
  (2014) 046, [\href{http://arxiv.org/abs/1401.4458}{{\tt arXiv:1401.4458}}].

\bibitem{Procura:2014cba}
M.~Procura, W.~J. Waalewijn, and L.~Zeune, {\it {Resummation of
  Double-Differential Cross Sections and Fully-Unintegrated Parton Distribution
  Functions}},  {\em JHEP} {\bf 02} (2015) 117,
  [\href{http://arxiv.org/abs/1410.6483}{{\tt arXiv:1410.6483}}].

\bibitem{Larkoski:2015uaa}
A.~J. Larkoski and I.~Moult, {\it {The Singular Behavior of Jet Substructure
  Observables}},  {\em Phys. Rev.} {\bf D93} (2016) 014017,
  [\href{http://arxiv.org/abs/1510.08459}{{\tt arXiv:1510.08459}}].

\bibitem{Salam:2016yht}
G.~P. Salam, L.~Schunk, and G.~Soyez, {\it {Dichroic subjettiness ratios to
  distinguish colour flows in boosted boson tagging}},
  \href{http://arxiv.org/abs/1612.03917}{{\tt arXiv:1612.03917}}.

\bibitem{Gallicchio:2012ez}
J.~Gallicchio and M.~D. Schwartz, {\it {Quark and Gluon Jet Substructure}},
  {\em JHEP} {\bf 04} (2013) 090, [\href{http://arxiv.org/abs/1211.7038}{{\tt
  arXiv:1211.7038}}].

\bibitem{Aad:2014gea}
{\bf ATLAS} Collaboration, G.~Aad et~al., {\it {Light-quark and gluon jet
  discrimination in $pp$ collisions at $\sqrt{s}=7\mathrm {\ TeV}$ with the
  ATLAS detector}},  {\em Eur. Phys. J.} {\bf C74} (2014), no.~8 3023,
  [\href{http://arxiv.org/abs/1405.6583}{{\tt arXiv:1405.6583}}].

\bibitem{Larkoski:2014pca}
A.~J. Larkoski, J.~Thaler, and W.~J. Waalewijn, {\it {Gaining (Mutual)
  Information about Quark/Gluon Discrimination}},  {\em JHEP} {\bf 11} (2014)
  129, [\href{http://arxiv.org/abs/1408.3122}{{\tt arXiv:1408.3122}}].

\bibitem{Izaguirre:2014ira}
E.~Izaguirre, B.~Shuve, and I.~Yavin, {\it {Improving Identification of Dijet
  Resonances at Hadron Colliders}},  {\em Phys. Rev. Lett.} {\bf 114} (2015),
  no.~4 041802, [\href{http://arxiv.org/abs/1407.7037}{{\tt arXiv:1407.7037}}].

\bibitem{Larkoski:2015kga}
A.~J. Larkoski, I.~Moult, and D.~Neill, {\it {Analytic Boosted Boson
  Discrimination}},  {\em JHEP} {\bf 05} (2016) 117,
  [\href{http://arxiv.org/abs/1507.03018}{{\tt arXiv:1507.03018}}].

\bibitem{Badger:2016bpw}
J.~R. Andersen et~al., {\it {Les Houches 2015: Physics at TeV Colliders
  Standard Model Working Group Report}},  in {\em {9th Les Houches Workshop on
  Physics at TeV Colliders (PhysTeV 2015) Les Houches, France, June 1-19,
  2015}}, 2016.
\newblock \href{http://arxiv.org/abs/1605.04692}{{\tt arXiv:1605.04692}}.

\bibitem{Aad:2016oit}
{\bf ATLAS} Collaboration, G.~Aad et~al., {\it {Measurement of the
  charged-particle multiplicity inside jets from $\sqrt{s}=8$ TeV $pp$
  collisions with the ATLAS detector}},  {\em Eur. Phys. J.} {\bf C76} (2016),
  no.~6 322, [\href{http://arxiv.org/abs/1602.00988}{{\tt arXiv:1602.00988}}].

\bibitem{Gras:2017jty}
P.~Gras, S.~Hoeche, D.~Kar, A.~Larkoski, L.~L�nnblad, S.~Pl�tzer,
  A.~Si�dmok, P.~Skands, G.~Soyez, and J.~Thaler, {\it {Systematics of
  quark/gluon tagging}},  \href{http://arxiv.org/abs/1704.03878}{{\tt
  arXiv:1704.03878}}.

\bibitem{Hocker:2007ht}
A.~Hocker et~al., {\it {TMVA - Toolkit for Multivariate Data Analysis}},  {\em
  PoS} {\bf ACAT} (2007) 040, [\href{http://arxiv.org/abs/physics/0703039}{{\tt
  physics/0703039}}].

\bibitem{Speckmayer:2010zz}
P.~Speckmayer, A.~Hocker, J.~Stelzer, and H.~Voss, {\it {The toolkit for
  multivariate data analysis, TMVA 4}},  {\em J. Phys. Conf. Ser.} {\bf 219}
  (2010) 032057.

\end{thebibliography}\endgroup
\end{document}